\documentclass[a4paper,11pt]{article}
\pdfoutput=1

 \usepackage{jheppub}
 \usepackage[T1]{fontenc}
 \usepackage[utf8]{inputenc}
 \usepackage{booktabs}
 \usepackage{xcolor}
 \usepackage{xspace}
 \usepackage{dcolumn}
 \usepackage{caption}
 \usepackage
 [subrefformat=parens]
 {subcaption}
 \usepackage[group-digits=false]{siunitx}
 \usepackage{microtype}
 \usepackage{lmodern}
 \usepackage{hyperref}

\newcommand{\Pl}{\ell}
\newcommand{\fb}{{\ensuremath\unskip\,\text{fb}}\xspace}
\def\refeq#1{\mbox{(\ref{#1})}}
\def\reffi#1{\mbox{Figure~\ref{#1}}}
\def\reffis#1{\mbox{Figures~\ref{#1}}}
\def\refta#1{\mbox{Table~\ref{#1}}}

\def\refse#1{\mbox{Section~\ref{#1}}}

\def\citere#1{\mbox{Ref.~\cite{#1}}}
\def\citeres#1{\mbox{Refs.~\cite{#1}}}

\newcommand{\ie}{\emph{i.e.}\ }

\def\be{\begin{equation}}
\def\ee{\end{equation}}

\newcommand{\PH}{\ensuremath{\text{H}}\xspace}
\newcommand{\Pj}{\ensuremath{\text{j}}\xspace}
\newcommand{\Pp}{\ensuremath{\text{p}}\xspace}
\newcommand{\Pe}{\ensuremath{\text{e}}\xspace}

\newcommand{\Pq}{\ensuremath{\text{q}}\xspace}
\newcommand{\Pt}{\ensuremath{\text{t}}\xspace}

\newcommand{\Pg}{\ensuremath{\text{g}}\xspace}

\newcommand{\PW}{\ensuremath{\text{W}}\xspace}
\newcommand{\PZ}{\ensuremath{\text{Z}}\xspace}

\newcommand{\Mt}{\ensuremath{m_\Pt}\xspace}

\newcommand{\MWOS}{\ensuremath{M_\PW^\text{OS}}\xspace}
\newcommand{\MW}{\ensuremath{M_\PW}\xspace}
\newcommand{\MZOS}{\ensuremath{M_\PZ^\text{OS}}\xspace}
\newcommand{\MZ}{\ensuremath{M_\PZ}\xspace}

\newcommand{\Gt}{\ensuremath{\Gamma_\Pt}\xspace}
\newcommand{\GH}{\ensuremath{\Gamma_\PH}\xspace}

\newcommand{\GZOS}{\ensuremath{\Gamma_\PZ^\text{OS}}\xspace}

\newcommand{\GWOS}{\ensuremath{\Gamma_\PW^\text{OS}}\xspace}

\newcommand{\GeV}{\ensuremath{\,\text{GeV}}\xspace}
\newcommand{\TeV}{\ensuremath{\,\text{TeV}}\xspace}

\newcommand{\alphas}{\ensuremath{\alpha_\text{s}}\xspace}

\newcommand{\GF}{\ensuremath{G_\mu}}

\newcommand{\ptsub}[1]{\ensuremath{p_{\text{T},#1}}\xspace}

\newcommand{\MVOS}{\ensuremath{M_{V}^\text{OS}}\xspace}%
\newcommand{\GVOS}{\ensuremath{\Gamma_{V}^\text{OS}}\xspace}%

\newcommand{\rd}{\mathrm{d}}
\newcommand{\cw}{c_{\mathrm{w}}}
\newcommand{\sw}{s_{\mathrm{w}}}

\newcommand{\newc}{\newcommand}
\newc{\bi}{\begin{itemize}}
\newc{\ei}{\end{itemize}}
\newc{\benu}{\begin{enumerate}}
\newc{\eenu}{\end{enumerate}}
\newc{\bc}{\begin{center}}
\newc{\ec}{\end{center}}
\newc{\bfig}{\begin{figure}}
\newc{\efig}{\end{figure}}
\newc{\qbar}{\bar{q}}
\newc{\go}{\tilde{g}}
\newc{\PB}{\textsc{Powheg-Box}}

\newcommand{\rT}{{\mathrm{T}}}
\newcolumntype{.}{D{.}{.}{-1}}
\newcolumntype{d}[1]{D{.}{.}{#1}}

\renewcommand{\vec}[1]{\mathbf{#1}}
\colorlet{tableoverheadcolor}{gray!37.5}
\colorlet{tableheadcolor}{gray!25}
\colorlet{tablerowcolor}{gray!12.5}

\newlength{\width}
\newlength{\height}

\def\draftdate{\relax}
\def\mda{\relax}
\def\mua{\relax}
\def\mla{\relax}
\def\draft{
\def\thtystars{******************************}
\def\sixtystars{\thtystars\thtystars}
\typeout{}
\typeout{\sixtystars**}
\typeout{* Draft mode!
         For final version remove \protect\draft\space in source file *}
\typeout{\sixtystars**}
\typeout{}
\def\draftdate{\today}
\def\mua{\marginpar[\boldmath\hfil$\uparrow$]%
                   {\boldmath$\uparrow$\hfil}\color{black}%
                    \typeout{marginpar: $\uparrow$}\ignorespaces}
\def\mda{\color{red}\marginpar[\boldmath\hfil$\downarrow$]%
                   {\boldmath$\downarrow$\hfil}%
                    \typeout{marginpar: $\downarrow$}\ignorespaces}
\def\mla{\marginpar[\boldmath\hfil$\rightarrow$]%
                   {\boldmath$\leftarrow $\hfil}%
                    \typeout{marginpar: $\leftrightarrow$}\ignorespaces}
\def\Mua{\marginpar[\boldmath\hfil$\Uparrow$]%
                   {\boldmath$\Uparrow$\hfil}\color{black}%
                    \typeout{marginpar: $\uparrow$}\ignorespaces}
\def\Mda{\color{red}\marginpar[\boldmath\hfil$\Downarrow$]%
                   {\boldmath$\Downarrow$\hfil}%
                    \typeout{marginpar: $\downarrow$}\ignorespaces}
\def\Mla{\marginpar[\boldmath\hfil\textcolor{red}{$\Rightarrow$}]%
                   {\boldmath\textcolor{red}{$\Leftarrow $}\hfil}%
                    \typeout{marginpar: $\leftrightarrow$}\ignorespaces}
\overfullrule 5pt
\oddsidemargin 15mm
\marginparwidth 29mm
}

\title{\strut\\[-42mm]
\strut\hfill\mbox{\small {FR-PHENO-2019-001}}\\[-.7ex]
\strut\hfill\mbox{\small {Cavendish-HEP-19/05}}\\[-.7ex]
\strut\hfill\mbox{\small {TIF-UNIMI-2019-1}}\\[-.7ex]
\strut\hfill\mbox{\small {VBSCAN-PUB-02-19}}
\\[1cm]
\vspace{13mm}
QCD and electroweak corrections to WZ scattering at the LHC}

\subheader{\today}

\author{Ansgar Denner$^1$,}
\author{Stefan Dittmaier$^2$,}
\author{Philipp Maierh\"{o}fer$^2$,}
\author{Mathieu Pellen$^3$,}
\author{Christopher Schwan$^{2,4}$}

\affiliation{$^1$Universit\"at W\"urzburg, %
        Institut f\"ur Theoretische Physik und Astrophysik, \\ %
        Emil-Hilb-Weg 22,  %
        97074 W\"urzburg, %
        Germany%
}
\affiliation{$^2$Universit\"at Freiburg,
        Physikalisches Insitut,
        Hermann-Herder-Stra\ss{}e 3,
        79104 Freiburg,
        Germany%
}
\affiliation{$^3$University of Cambridge, Cavendish Laboratory,
        Cambridge CB3 0HE, United Kingdom%
}
\affiliation{$^4$Tif Lab, Dipartimento di Fisica, Universit\`a di Milano and INFN,
    Sezione di Milano,
    Via Celoria 16,
    20133 Milano,
    Italy
}
\emailAdd{ansgar.denner@physik.uni-wuerzburg.de}
\emailAdd{stefan.dittmaier@physik.uni-freiburg.de}
\emailAdd{philipp.maierhoefer@physik.uni-freiburg.de}
\emailAdd{mpellen@hep.phy.cam.ac.uk}
\emailAdd{christopher.schwan@mi.infn.it}

\abstract{We present the first computation of the full
  next-to-leading-order QCD and electroweak corrections to the WZ
  scattering process at the LHC.  All off-shell, gauge-boson-decay,
  and interference effects are taken into account for the process
  $\Pp\Pp \to \mu^+\mu^-\Pe^+\nu_\Pe\Pj\Pj+X$ at the orders
  $\mathcal{O}{\left( \alphas \alpha^6 \right)}$ and
  $\mathcal{O}{\left( \alpha^7 \right)}$.  
  The electroweak corrections feature the typical Sudakov
  behaviour towards high energy and amount to
  $-16\%$ relative to the electroweak contribution to the 
  integrated cross section. Moreover, the corrections induce
  significant shape distortions in differential distributions.
  The next-to-leading-order analysis of the quark- and gluon-induced channels is
  supplemented by a leading-order study of all possible contributions
  to the full $4\ell+2\mbox{jets}$ production cross section
  in a realistic fiducial phase-space volume.
}

\begin{document}

\maketitle

\newpage

\section{Introduction}

The accumulation of experimental data during Run~II of the Large
Hadron Collider (LHC) allows to measure some rare Standard Model (SM)
processes for the first time.  Vector-boson scattering (VBS) processes
constitute a prime example of processes that have not been measured
before Run~II.  While the scattering of like-sign $\PW$-boson pairs,
the golden VBS channel, has been measured
first
\cite{Aad:2014zda,Khachatryan:2014sta,Aaboud:2016ffv,Sirunyan:2017ret,ATLAS-CONF-2018-030},
the $\PW\PZ$ channel comes in second \cite{ATLAS:2018ucv,Sirunyan:2019ksz}.
It features a lower cross section than $\PW^\pm\PW^\pm$ scattering,
but has only one neutrino in the final state, allowing thus for better
reconstruction and a better study of its properties.

As experimental errors (both statistical and systematic) will shrink
in the next few years, precise theoretical predictions should be
carefully prepared.  In particular, higher-order corrections of both
QCD and electroweak (EW) type should be incorporated.  The inclusion
of next-to-leading-order (NLO) QCD corrections has become standard for
LHC analyses, but 
not yet the inclusion of 
EW corrections, which are known to increase at high
energies owing to Sudakov logarithms.  For the class of VBS processes,
EW corrections are expected to be particularly large
\cite{Biedermann:2016yds}.  This expectation was confirmed in the
first complete NLO QCD+EW calculation presented in
\citere{Biedermann:2017bss} for like-sign $\PW\PW$~scattering where it
turned out that the genuine EW corrections of order
$\mathcal{O}{\left( \alpha^7 \right)}$ are even the largest NLO
contribution.

In this article, we present results for the first calculation of the
full NLO QCD+EW corrections to the $\PW\PZ$ scattering process 
at the LHC
with the final state $\mu^+\mu^-\Pe^+\nu_\Pe\Pj\Pj$.
An analysis of the LO contributions to the
$\PW\PZ\Pj\Pj$ production mode was presented in
\citere{Bendavid:2018nar}, where also different Monte Carlo programs were
compared.
In our NLO analysis, we include the whole set of
contributing diagrams in the relevant orders, instead of only VBS
configurations.
The QCD and
especially the EW corrections are rather involved, as the process
features seven charged external particles.  This is the first time that
EW corrections are computed for a process involving so many charged
particles.
The leptonic final state with a single net charge gives rise to a larger number of partonic channels as compared to like-sign WW VBS, which complicates the calculation further.
We also note that our calculation of QCD corrections is based on the full set of NLO
diagrams including all interferences without approximation, \emph{i.e.}\ we
do not employ the so-called VBS approximation used in previous QCD
calculations~\cite{Bozzi:2007ur,Jager:2018cyo}, which neglects colour exchange
between the two incoming protons.  While for the current experimental
precision such approximations are most likely sufficient, in the
future they might actually be inadequate, because they can fail at
the level of $10\%$ in differential distributions, as shown in
\citere{Ballestrero:2018anz} for like-sign $\PW\PW$~scattering.

In addition to the contributions to the NLO cross section
of orders $\mathcal{O}{\left( \alphas \alpha^6 \right)}$ and $\mathcal{O}{\left( \alpha^7 \right)}$, we also provide predictions for all LO processes relevant for the $\mu^+\mu^-\Pe^+\nu_\Pe\Pj\Pj$ final state.
These include the orders $\mathcal{O}{\left( \alpha^6 \right)}$ (EW contribution), $\mathcal{O}{\left( \alphas \alpha^5 \right)}$ (interference), and $\mathcal{O}{\left( \alphas^2 \alpha^4 \right)}$ (QCD contribution).
Contributions including photons in the initial state or external
bottom quarks
are discussed separately.

All these results are presented in the form of cross sections and
differential distributions for realistic experimental cuts.
Specifically, the event selection chosen is the so-called \emph{loose
  fiducial} region presented by the CMS collaboration in
\citere{Sirunyan:2019ksz}.  It has the advantage to be simple enough to
be implemented easily in a Monte Carlo program.  Such an experimental
effort is particularly welcome 
by theorists as it allows for a direct use of
state-of-the-art theoretical predictions in experimental analyses.

Finally, we would like to mention that all results have been produced by two independent Monte Carlo programs, matrix elements providers, and loop libraries: 
One is the Monte Carlo program {\sc BONSAY} with matrix elements from {\sc OpenLoops}~\cite{Cascioli:2011va,Kallweit:2014xda} and loop integrals evaluated with the {\sc DD} mode of the {\sc Collier}~\cite{Denner:2014gla,Denner:2016kdg} library.
The other Monte Carlo is {\sc MoCaNLO} with matrix elements from {\sc Recola}~\cite{Actis:2012qn,Actis:2016mpe} and loop integrals evaluated with the {\sc COLI} mode of the {\sc Collier} library.
The two independent calculations ensure a exhaustive validation of all results presented in this paper.

This article is organised as follows:
In \refse{sec:process}, the various contributions
to the NLO cross section of the considered process are described.
In \refse{sec:details} the details on the implementations of the
computation as well as the checks performed to validate the results
are presented.  Section~\ref{sec:results} is devoted to the
description and the analysis of the results.  Finally,
\refse{sec:conclusion} contains a summary of the article as well as
concluding remarks.

\section{Definition of the process and survey of cross-section contributions}
\label{sec:process}

\subsection{Leading-order contributions}

\begin{figure}
\centering
\begin{subfigure}{0.33\textwidth}
\centering
\includegraphics{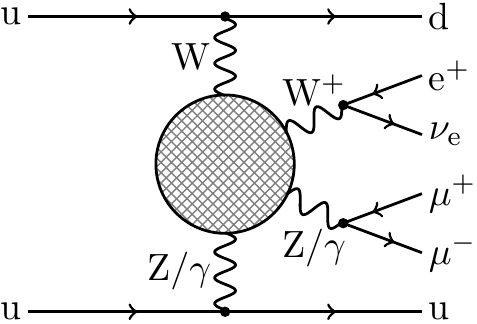}
\caption{VBS, doubly-resonant}
\label{fig:born_qq_vbs}
\end{subfigure}
\begin{subfigure}{0.33\textwidth}
\centering
\includegraphics{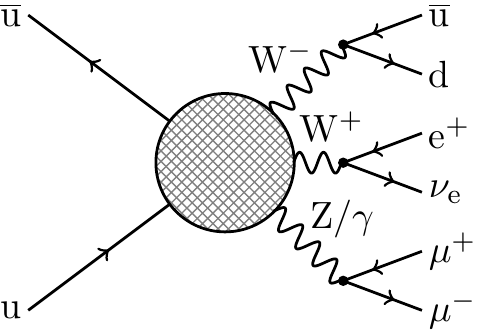}
\caption{triple-boson production}
\label{fig:born_qq_WWZ}
\end{subfigure}%
\begin{subfigure}{0.33\textwidth}
\centering
\includegraphics{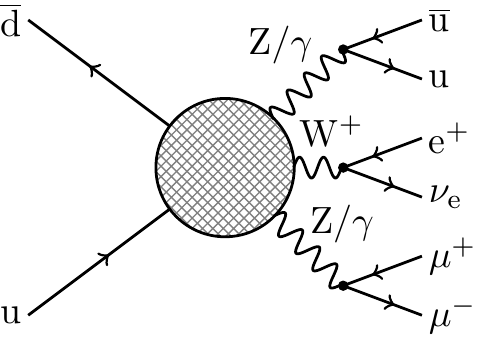}
\caption{triple-boson production}
\label{fig:born_qq_WZZ}
\end{subfigure}%
\par\bigskip
\begin{subfigure}{0.33\textwidth}
\centering
\includegraphics{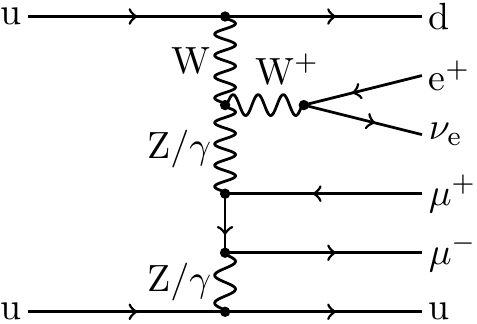}
\caption{singly-resonant}
\label{fig:born_qq_W}
\end{subfigure}%
\begin{subfigure}{0.33\textwidth}
\centering
\includegraphics{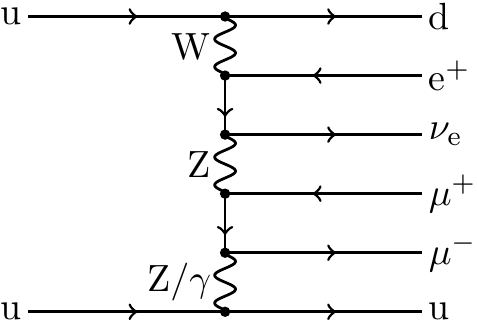}
\caption{non-resonant}
\label{fig:born_qq_nonres}
\end{subfigure}%
\begin{subfigure}{0.33\textwidth}
\centering
\includegraphics{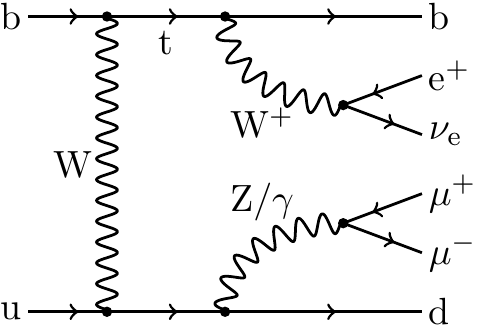}
\caption{top-Z-jet production ($\mathrm{t}\mathrm{Z}\mathrm{j}$)}
\label{fig:born_bq_tZj}
\end{subfigure}\par\bigskip
\begin{subfigure}{0.33\textwidth}
\centering
\includegraphics{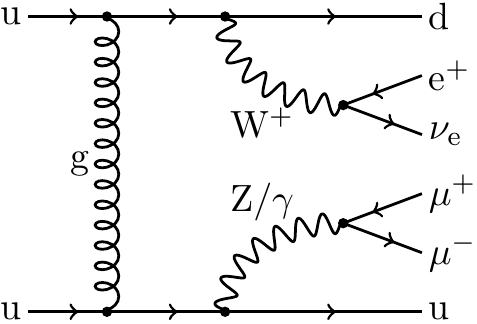}
\caption{QCD contribution}
\label{fig:born_qq_gluon}
\end{subfigure}%
\begin{subfigure}{0.33\textwidth}
\centering
\includegraphics{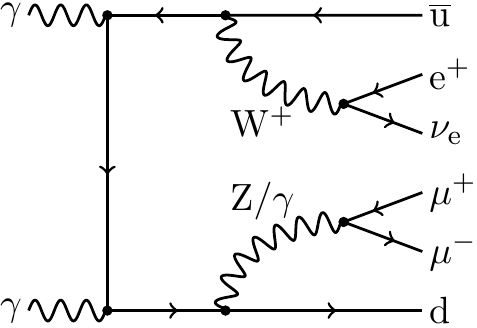}
\caption{photon--photon initiated}
\label{fig:born_gaga}
\end{subfigure}%
\begin{subfigure}{0.33\textwidth}
\centering
\includegraphics{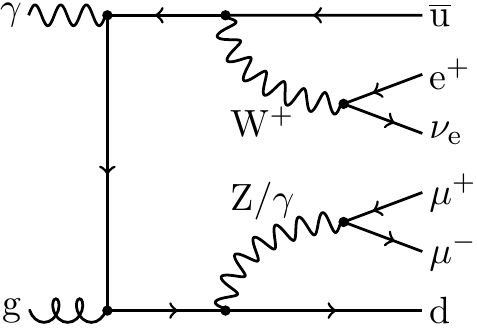}
\caption{photon--gluon initiated}
\label{fig:born_gagl}
\end{subfigure}%
\par\bigskip
\begin{subfigure}{0.33\textwidth}
\centering
\includegraphics{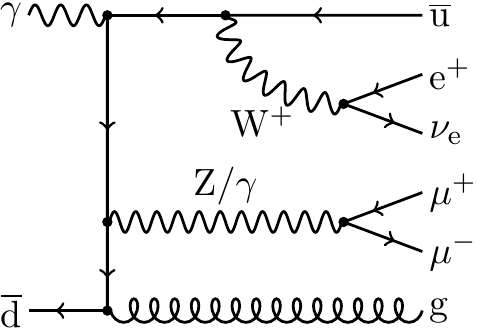}
\caption{photon--quark initiated}
\label{fig:born_qga}
\end{subfigure}%
\begin{subfigure}{0.33\textwidth}
\centering
\includegraphics{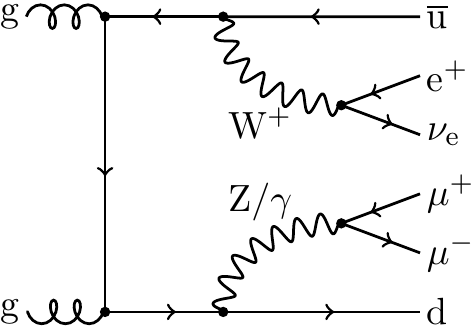}
\caption{gluon--gluon initiated}
\label{fig:born_glgl}
\end{subfigure}
\begin{subfigure}{0.33\textwidth}
\centering
\includegraphics{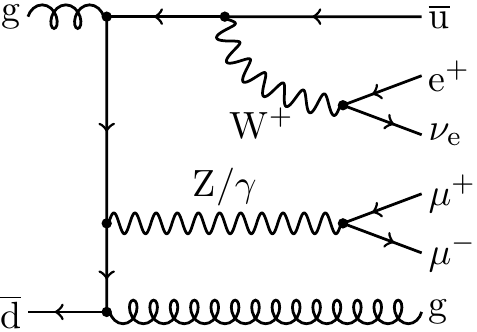}
\caption{gluon--quark initiated}
\label{fig:born_qgl}
\end{subfigure}%
\caption{Examples of LO Feynman diagrams.}
\label{fig:borndiagrams}
\end{figure}
As for all quark--quark-initiated processes characterised by four
leptons and two jets in the final state, two types of amplitudes occur
for the quark-induced processes $\Pq\Pq \to
\mu^+\mu^-\Pe^+\nu_\Pe\Pq\Pq$, 
where $\Pq$ generically stands for a
quark or antiquark: 
These are diagrams of order $\mathcal{O}{\left(
    g^6 \right)}$ (some sample diagrams are shown in 
\reffis{fig:born_qq_vbs}--\ref{fig:born_bq_tZj}) and diagrams of 
order $\mathcal{O}{\left( g_{\rm s}^2 g^4 \right)}$ (an example is
depicted in \reffi{fig:born_qq_gluon}), with $g_{\rm s}$ and $g$ generically
denoting the strong and electroweak gauge couplings, respectively.
Besides VBS the former diagrams involve also the production of three vector bosons as well as singly-resonant and non-resonant diagrams.
Consequently, three different orders contribute to the LO cross
section: $\mathcal{O}{\left( \alpha^6 \right)}$ (EW contribution),
$\mathcal{O}{\left( \alphas \alpha^5 \right)}$ (interference), and
$\mathcal{O}{\left( \alphas^2 \alpha^4 \right)}$ (QCD contribution).

At the order $\mathcal{O}{\left( \alpha^6 \right)}$, where all
couplings are of EW origin, there are in addition
contributions from $\gamma \gamma \to \mu^+\mu^-\Pe^+\nu_\Pe\Pq\Pq$
(\reffi{fig:born_gaga} provides an example).  For the quark
contributions, one can further distinguish the cases in which an
external quark is a bottom quark or a light one.
In our predictions, we show separately the $\mathcal{O}{\left( \alpha^6 \right)}$ contributions with an external bottom quark, as the corresponding partonic channels can develop a top-quark resonance (see \reffi{fig:born_bq_tZj}).
The photon-induced contributions are also shown separately.

The order $\mathcal{O}{\left( \alphas \alpha^5 \right)}$ contributions
are obtained by interfering amplitudes of $\mathcal{O}{\left( g^6
\right)}$ and $\mathcal{O}{\left( g_{\rm s}^2 g^4 \right)}$ in the
channels $\Pq \Pq \to \mu^+\mu^-\Pe^+\nu_\Pe\Pq\Pq$.  Further
contributions in this order result from squares of amplitudes of order
$\mathcal{O}{\left( g_{\rm s} g^5 \right)}$ of the channels $\Pg \gamma
\to \mu^+\mu^-\Pe^+\nu_\Pe\Pq\Pq$ and $\Pq \gamma \to
\mu^+\mu^-\Pe^+\nu_\Pe\Pq\Pg$ (see \reffis{fig:born_gagl} and
\ref{fig:born_qga} for examples).  Those contributions
are shown together with the order $\mathcal{O} (\alpha^6)$ photon-induced contributions.

Finally, the $\mathcal{O}{\left( \alphas^2 \alpha^4 \right)}$
contributions result from channels with either four external quarks or
two external quarks and two gluons (see \reffis{fig:born_glgl} and
\ref{fig:born_qgl} for examples).  The contribution with two gluons
in the initial state is particularly large due to the large gluonic
parton-distribution function (PDF) at the LHC.
The order $\mathcal{O} (\alpha_\mathrm{s}^2 \alpha^4)$ contributions with external bottom quarks are shown separately, in combination with the order $\mathcal{O} (\alpha^6)$ bottom-quark contributions.

\subsection{Virtual corrections}

\begin{figure}
\centering
\begin{subfigure}{0.33\textwidth}
\centering
\includegraphics{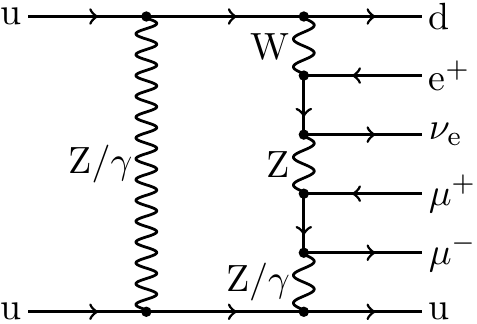}
\caption{8-point function}
\label{fig:virt_8pt}
\end{subfigure}%
\begin{subfigure}{0.33\textwidth}
\centering
\includegraphics{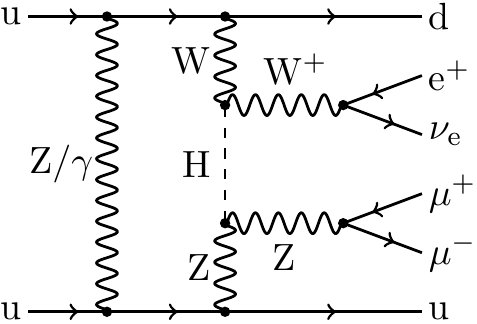}
\caption{6-point function with Higgs}
\label{fig:virt_Higgs}
\end{subfigure}%
\begin{subfigure}{0.33\textwidth}
\centering
\includegraphics{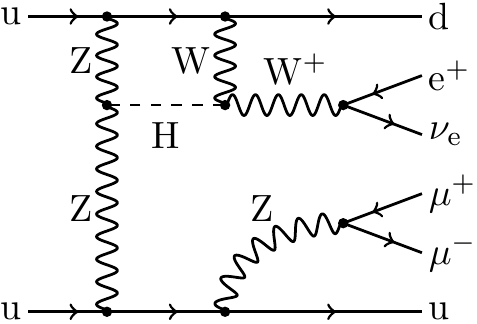}
\caption{Higgs exchange}
\label{fig:virt_Higgsb}
\end{subfigure}\par\bigskip
\begin{subfigure}{0.33\textwidth}
\centering
\includegraphics{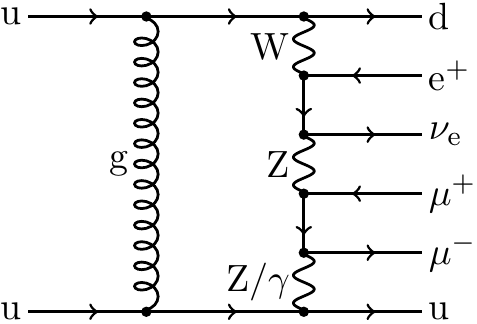}
\caption{gluon between quark lines}
\label{fig:virt_gl}
\end{subfigure}%
\begin{subfigure}{0.33\textwidth}
\centering
\includegraphics{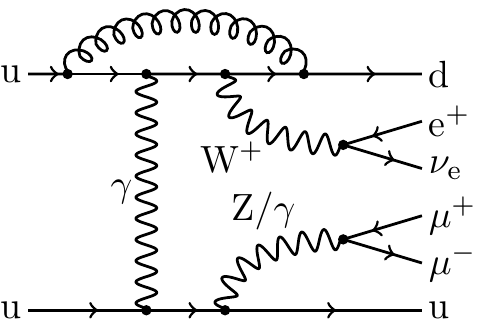}
\caption{gluon at single quark line}
\label{fig:virt_gl_sameline}
\end{subfigure}%
\caption{Example loop diagrams.}
\end{figure}

We compute the NLO corrections of orders $\mathcal{O}{(\alpha^7)}$ and
$\mathcal{O}{(\alphas\alpha^6)}$ for the process
\begin{equation}
  \Pp \Pp \to \mu^+\mu^-\Pe^+\nu_\Pe\Pj\Pj + X.
\end{equation}
Virtual corrections of order $\mathcal{O}{(\alpha^7)}$ result from interferences
of the tree-level EW diagrams of order $\mathcal{O}{(g^6)}$ with purely EW loop
diagrams of order $\mathcal{O}{(g^8)}$.
Examples for the latter are depicted in
\reffis{fig:virt_8pt}--\ref{fig:virt_Higgsb}.
The virtual corrections of order $\mathcal{O}{(\alphas\alpha^6)}$ receive
contributions from several sources.
EW loop diagrams for quark-induced processes of order $\mathcal{O}{(g^8)}$
(\reffis{fig:virt_8pt}--\ref{fig:virt_Higgsb}) interfere with LO diagrams of
order $\mathcal{O}{(g_{\mathrm{s}}^2g^4)}$ (\reffi{fig:born_qq_gluon}).
Due to the SU(3) colour structure, this only gives a non-vanishing
contribution for partonic processes where all external quarks belong to the same
generation.
Loop diagrams of order $\mathcal{O}{(g_{\mathrm{s}}^2g^6)}$ (like in
\reffis{fig:virt_gl}--\ref{fig:virt_gl_sameline}) interfere with EW LO diagrams.
Owing to the colour structure, in case of two different generations of quarks in
the partonic process, only diagrams of the type \reffi{fig:virt_gl_sameline}
with gluon exchange within one quark line contribute.
In both types of NLO corrections, partonic channels with initial-state photons are not taken into account, since their contribution is already strongly suppressed at LO.
Channels with external bottom quarks are excluded as well.
Those could only significantly contribute via singly-resonant top quarks, which corresponds
to a different experimental signature.
In total, $40$ partonic channels must be taken into account at each coupling
order with up to \mbox{$\sim 83{,}000$} 1-loop Feynman diagrams contributing per
channel.
Tensor integrals appear up to $8$-point functions with tensor ranks of up to $4$.

In the VBS approximation,
as employed in previous QCD
calculations, only QCD corrections of the type \reffi{fig:virt_gl_sameline}
with gluon exchange within one quark line are taken into account.
With \mbox{$\sim 1000$} Feynman diagrams per partonic channel and up to 5-point
functions this approximation reduces the complexity drastically in comparison to the calculation presented in this article.

\subsection{Real corrections}

\begin{figure}
\centering
\begin{subfigure}{0.33\textwidth}
\centering
\includegraphics{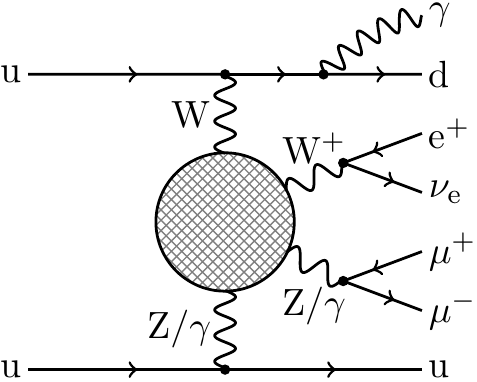}
\caption{photon emission, VBS}
\label{fig:real-ga_vbs}
\end{subfigure}%
\begin{subfigure}{0.33\textwidth}
\centering
\includegraphics{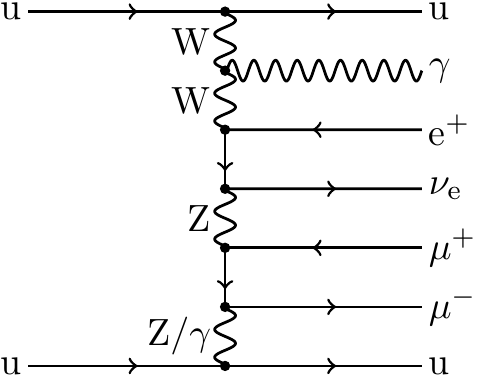}
\caption{non-resonant}
\label{fig:real-ga_nonres}
\end{subfigure}%
\begin{subfigure}{0.33\textwidth}
\centering
\includegraphics{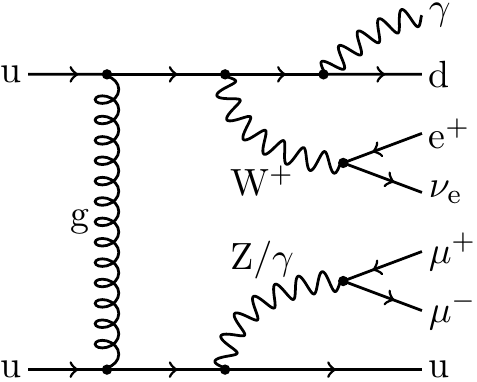}
\caption{QCD production}
\label{fig:real-ph_vbs}
\end{subfigure}\par\bigskip
\begin{subfigure}{0.33\textwidth}
\centering
\includegraphics{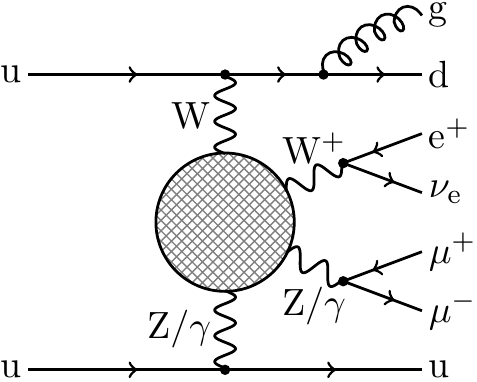}
\caption{gluon emission, VBS}
\label{fig:real-gl_vbs}
\end{subfigure}%
\caption{Sample diagrams for real corrections.}
\end{figure}
At the order $\mathcal{O}{\left( \alpha^7 \right)}$, 
there are two types of real EW corrections:
One is due to photon radiation, which results from
radiating a photon from one of the charged particles of the LO
processes of order $\mathcal{O}{\left( \alpha^6 \right)}$.  
The other type comprises photon-induced channels, which we do
not take into account at NLO, as already the
corresponding LO contribution turned out to be very small.\footnote{In the case of same-sign W scattering, photon-induced corrections of order $\mathrm{O}(\alpha^7)$ have been found to be below \SI{2}{\percent}~\cite{Biedermann:2017bss}.
They are expected to be of similar size for WZ~scattering.} 
Therefore, only
real photon radiation from the $\Pq \Pq \to
\mu^+\mu^-\Pe^+\nu_\Pe\Pq\Pq$ channels, \ie the process  $\Pq \Pq \to
\mu^+\mu^-\Pe^+\nu_\Pe\Pq\Pq \gamma$, is considered.  Some
relevant Feynman diagrams are shown in \reffis{fig:real-ga_vbs}
and \ref{fig:real-ga_nonres}. The related infrared (IR) divergences
are subtracted using QED dipole subtraction~\cite{Dittmaier:1999mb,Dittmaier:2008md}.

At the order $\mathcal{O}{\left( \alphas \alpha^6 \right)}$ 
a mixture of two
types of real radiation contributes, because this NLO contribution
comprises both QCD corrections to the order $\mathcal{O}{\left(
  \alpha^6 \right)}$ and EW corrections to the order $\mathcal{O}
{\left( \alphas \alpha^5 \right)}$.  The EW corrections are obtained 
by attaching a photon to each LO diagram of
order $\mathcal{O}{\left( g_{\rm s}^2 g^4 \right)}$ in all possible ways
(see \reffi{fig:real-ph_vbs}) and interfering the resulting diagrams with
all photon emission diagrams of $\mathcal{O}{\left( g^7 \right)}$.
The QCD
corrections are obtained by attaching a gluon to each LO diagram of
order $\mathcal{O}{\left( g^6 \right)}$ in all possible ways (a
sample diagram is given in \reffi{fig:real-gl_vbs}) resulting in the
process $\Pq \Pq \to \mu^+\mu^-\Pe^+\nu_\Pe\Pq\Pq \Pg$, and
squaring the corresponding amplitude.
Of course, there are also real QCD corrections of the same order with the 
gluon crossed into the initial state, $\Pg \Pq/\Pq \Pg \to \mu^+\mu^-\Pe^+\nu_\Pe\Pq\Pq \Pq$.
The phase-space integration for the real corrections of
$\mathcal{O}{\left(\alphas \alpha^6 \right)}$ leads to both QCD and EW
IR divergences in the limits of soft and/or collinear gluon or
photon emission, 
or via forward branchings of QCD partons in the initial state. 
Figure~\ref{fig:real-ga_vbs} shows a $\Pq \to \Pq \gamma$ splitting
of QED type, \reffi{fig:real-gl_vbs} displays a typical $\Pq \to \Pq
\Pg$ splitting of QCD type.

\begin{figure}
\centering
\begin{subfigure}{0.33\textwidth}
\centering
\includegraphics{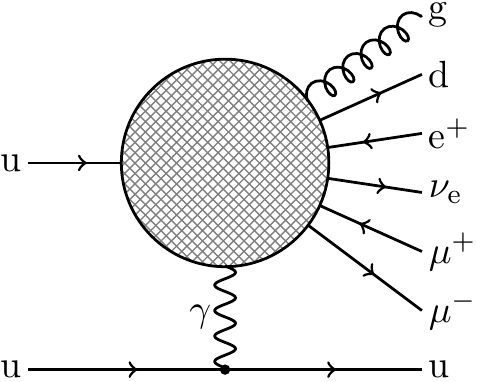}
\caption{initial-state coll.\ sing.}
\label{fig:real-gl_qgaq-split}
\end{subfigure}%
\begin{subfigure}{0.33\textwidth}
\centering
\includegraphics{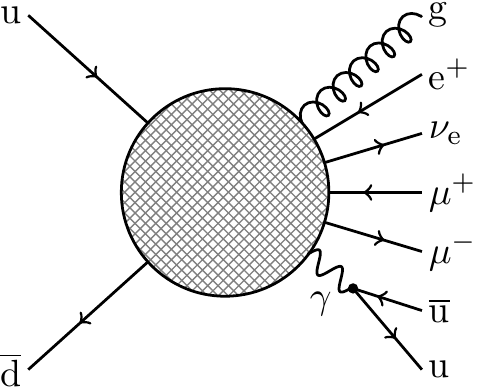}
\caption{final-state coll.\ sing.}
\label{fig:real-gl_gaqq-split}
\end{subfigure}%
\begin{subfigure}{0.33\textwidth}
\centering
\includegraphics{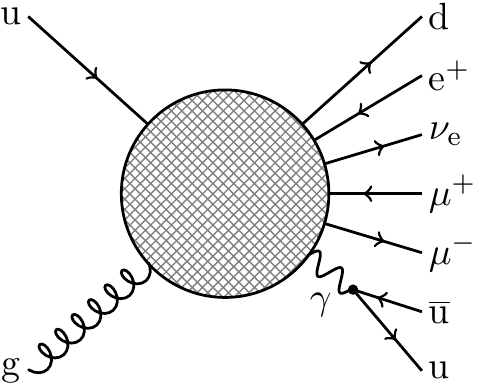}
\caption{final-state coll.\ sing.}
\label{fig:inc-gl_gaqq-split}
\end{subfigure}%
\caption{Photon singularities in the $\mathcal{O}{(\alpha_\mathrm{s} \alpha^6)}$ real corrections.}
\label{fig:photon-singularities}
\end{figure}

For real radiation of order $\mathcal{O}{\left( \alphas \alpha^6
  \right)}$ further subtleties arise.  Some diagrams with external real
gluons involve  singularities associated with soft/collinear photons.
One example is given in \reffi{fig:real-gl_qgaq-split} which has an
initial-state collinear singularity and requires both QED and QCD dipoles
to subtract all IR divergences. 
Another subtle case arises from final
states involving a $\Pq\bar\Pq$ pair.
This pair can result from a QED
splitting $\gamma^* \to \Pq\bar\Pq$ where the off-shell photon has a
very low virtuality (see \reffis{fig:real-gl_gaqq-split} and \ref{fig:inc-gl_gaqq-split}).   
In the singular limit where this virtuality goes to
zero, a collinear singularity develops with a universal singular
structure factorising from the hard matrix elements of the underlying
process with a real photon instead of the $\Pq\bar\Pq$ pair.  Note,
however, that the physical final state is still a jet, or at least
some hadronic activity, emerging from the photon initiating the
splitting.  Technically, the collinear singularity can, \emph{e.g.}, be
separated via dipole subtraction as described in
\citere{Dittmaier:2008md}, \emph{i.e.}\ an auxiliary \emph{subtraction function}
is subtracted from the original integrand, rendering the resulting
contribution integrable over the singular region.  The formerly
subtracted contribution is added back after integration over the
singular region with the help of some 
regularisation, either by
switching from four to $D$ space--time dimensions or by employing
small quark masses.  Either way, the resulting singular contribution
is not yet described in a physically meaningful way, since the
splitting contains non-perturbative contributions. In the case of
low-virtuality $\gamma^* \to \Pq\bar\Pq$ splittings, this contribution
can be obtained from a dispersion integral for the $R$ ratio of
the cross sections for $\Pe^+\Pe^-\to\mathrm{hadrons}/\mu^+\mu^-$, as
will be further detailed in \citere{xxx}.  As described there, this contribution can
be tied to the quantity $\Delta\alpha_{\mathrm{had}}$, which is
derived from experimental data.  In our calculation we follow this
approach, \emph{i.e.}\ we separate the singularity via dipole
subtraction~\cite{Dittmaier:2008md} and add the non-perturbative
fragmentation-like ``photon-to-jets conversion part'' from the collinear region based on
$\Delta\alpha_{\mathrm{had}}$.
Conceptually, it is quite important to properly treat this non-perturbative contribution,
but in the present case the overall contribution matters only at the level of
$10^{-4}$ relative to the EW LO cross section.

Note that the previous discussion also applies in principle to the real corrections of order $\mathcal{O} (\alpha^7)$ with an additional photon.
However, the extra collinear singularities coming from the matrix elements similar to the ones in Fig.~\ref{fig:photon-singularities}, but with external photons instead of gluons, are cut off due to our process definition.
In particular, we treat a final-state photon in the real matrix elements always as a photon and never as a jet (or photon-jet).
This choice implies that phase-space points with such collinear singularities have either zero or one jet.
For final-state singularities, the two collinear quarks are clustered into a single jet while for an initial-state singularity, the collinear quark is along the beam pipe making it undetectable.
Therefore our requirement of having at least two jets (see \refse{ssec:InputParameters}) cuts away such singularities, rendering the real corrections finite, so that no additional terms of type $\gamma \to q\bar q$ are required. 

\section{Details of the computation}
\label{sec:details}

\subsection{Implementations}

In order to ensure the correctness of the results, two independent
Monte Carlo programs have been developed based on two entirely different sets
of matrix elements constructed by independent matrix-element providers.
One calculation is based on the combination {\sc BONSAY+OpenLoops},
the other on {\sc MoCaNLO+Recola}.

The program {\sc BONSAY} is a general-purpose Monte Carlo integrator
which can be used to calculate arbitrary NLO EW, QCD, and mixed corrections with
matrix elements from an external provider.
It has already been used before in \citere{Ballestrero:2018anz}
to calculate QCD corrections of like-sign WW~scattering.
It employs many different phase-space mappings that are combined via
multi-channel techniques~\cite{hep-mc}, similar to the {\sc Lusifer}
Monte Carlo program~\cite{Dittmaier:2002ap}, but allows to run the
integration in parallel on clusters using MPI~\cite{MpiForum}.

{\sc MoCaNLO} is also a generic Monte Carlo program, designed to
compute arbitrary cross sections in the SM at NLO QCD and/or EW
accuracy.  The efficient integration is ensured by using phase-space
mappings similar to the ones of
\citeres{Denner:1999gp,Dittmaier:2002ap,Berends:1994pv}.  It has
already been used to compute NLO QCD and EW corrections for several 
high-multiplicity processes \cite{Denner:2015yca,Denner:2016jyo,Denner:2016wet,Denner:2017kzu}, including the like-sign W-boson VBS
process~\cite{Biedermann:2016yds,Biedermann:2017bss,Ballestrero:2018anz}.
Moreover, it has also been tested against other independent codes for
the computation of EW corrections to di-boson production in
\citere{Bendavid:2018nar}.

In both Monte Carlo programs, IR divergences in the real radiation are
handled with the dipole-subtraction method for QCD
\cite{Catani:1996vz} and its extension to QED
\cite{Dittmaier:1999mb,Dittmaier:2008md}.  Although the same
algorithms are used, the two implementations are completely
independent.  The library {\sc LHAPDF}~\cite{Buckley:2014ana} provides
PDFs in both codes.

Both {\sc OpenLoops}~\cite{Cascioli:2011va,Kallweit:2014xda} and
{\sc Recola}~\cite{Actis:2012qn,Actis:2016mpe} use the {\sc Collier}
library~\cite{Denner:2014gla,Denner:2016kdg} to obtain numerically
stable results for the one-loop scalar
\cite{'tHooft:1978xw,Beenakker:1988jr,Dittmaier:2003bc,Denner:2010tr}
and tensor integrals
\cite{Passarino:1978jh,Denner:2002ii,Denner:2005nn}.
In order to
ensure independence, the two different modes of {\sc Collier} have
been used: the {\sc DD} mode in {\sc BONSAY+OpenLoops} and the
{\sc COLI} mode in {\sc MoCaNLO+Recola}.
The intermediate W/Z-boson resonances are treated in the complex-mass scheme \cite{Denner:1999gp,Denner:2005fg,Denner:2006ic} to ensure gauge independence of all LO and NLO amplitudes.
In \textsc{OpenLoops}, we set the option \texttt{use\_cms=2} to switch to the
same conventions for the complex-mass scheme as used by \textsc{Recola}.
This allows us to verify the agreement of the results of the two matrix-element
providers for individual phase-space points.

The numerical results presented in this article are obtained from BONSAY,\footnote{The only exception is the results of Table \ref{tab:NLO14} for a centre-of-mass of $14\TeV$ which has been obtained from {\sc MoCaNLO}.}
which agrees with the other implementation within integration errors, which
are typically of the size of a per mille with respect to the LO prediction.

\subsection{Validation}
\label{ssec:validation}

The first and strongest validation is that the final results (at the level of cross sections and for each bin of the differential distributions) of the two calculations agree within statistical errors.
This constitutes a very solid check as the two Monte Carlo programs as well as the matrix-element providers are different and independent.
This ensures the correct implementation of the event selection, input parameters as well as the subtraction on the one hand.
On the other hand, it also ensures the validity of the matrix elements used.

The $\alpha_{\mathrm{dipole}}$ parameter~\cite{Nagy:1998bb} allows one to restrict the phase space
to the singular regions, where $\alpha_{\mathrm{dipole}}=1$ corresponds to the full
phase space (within the acceptance defined by selection cuts) without additional restrictions.
Varying $\alpha_{\mathrm{dipole}}$ allows then for a robust check of the subtraction procedure.
Representative contributions have been checked between {\sc BONSAY+OpenLoops} and {\sc MoCaNLO+Recola} for $\alpha_{\mathrm{dipole}} = 1$ for both orders $\mathcal{O}{\left( \alphas \alpha^6 \right)}$ and $\mathcal{O}{\left( \alpha^7 \right)}$.
The final results have been obtained with $\alpha_{\mathrm{dipole}} = 1$ for {\sc BONSAY+OpenLoops} and $\alpha_{\mathrm{dipole}} = 10^{-2}$ for {\sc MoCaNLO+Recola}, showing agreement
at the level of a per mille.
This constitutes a strong check on the subtraction procedure used.

In addition, point-wise comparisons of (squared) matrix-element contributions
have been carried out for the virtual corrections.
At the order $\mathcal{O}{\left( \alpha^7 \right)}$, which comprises the numerically
most delicate loop amplitudes,
for 1000 phase-space points chosen in the fiducial volume described above,
more than $99\%$ of the points show at least 6 digits of agreement.
In total, the level of agreement spans from about 6 to 12 digits.

Finally, 1000 points have been generated to check the real QCD corrections.
This ensures the correct implementation of the event selection for both the
real radiation and the dipoles.
In that way, the correct implementation of the dynamical scale is ensured as well.

\section{Numerical results}
\label{sec:results}

\subsection{Input parameters and event selection}
\label{ssec:InputParameters}

The results presented are for the LHC operating at a centre-of-mass (CM)
energy of $13\TeV$.
We use the
NLO NNPDF 3.1 QED set~\cite{Ball:2014uwa,Bertone:2017bme} with the photon PDF determined by the LUXqed method~\cite{Manohar:2016nzj,Manohar:2017eqh} and $\alphas(\MZ) = 0.118$ ({\sc LHAPDF} ID 324900),
employing the fixed $N_\text{F}=5$ flavour scheme throughout.
We use the same PDFs for LO and NLO predictions.
Both QCD and QED singularities from collinear initial-state splittings
are factorised into redefined PDFs using the
${\overline{\rm MS}}$ factorisation scheme.

The central renormalisation and factorisation scales,
$\mu_{\mathrm{ren}}$ and $\mu_{\mathrm{fact}}$, are set to the
geometric average of the transverse momenta of the jets,
\begin{equation}
\label{eq:defscale}
 \mu_0 = \sqrt{p_{\rm T, j_1}\, p_{\rm T, j_2}}.
\end{equation}
The choice of this scale is motivated by the results of
\citere{Denner:2012dz} on like-sign WW~scattering, where it was shown
that this choice reduces the difference between
LO and NLO QCD predictions at large transverse momenta significantly.
In the following, we perform a 7-point scale variation of
the renormalisation and factorisation scales, \emph{i.e.}\ apart from the
``diagonal'' variations
$\mu_{\mathrm{ren}}=\mu_{\mathrm{fact}}=\mu_0$, $\mu_0/2$, $2\mu_0$
we set each of the two scales to $\mu_0/2$, $2\mu_0$ while
keeping the other scale fixed.

Regarding the electromagnetic coupling, the $G_\mu$
scheme (see, \emph{e.g.}\ \citeres{Denner:2000bj,Dittmaier:2001ay}) is used,
\emph{i.e.}\ the coupling is obtained from the Fermi constant $G_\mu$ as
\begin{equation}
  \alpha = \frac{\sqrt{2}}{\pi} G_\mu \MW^2 \left( 1 - \frac{\MW^2}{\MZ^2} \right)  \qquad \text{with}  \qquad   {\GF    = 1.16638\times 10^{-5}\GeV^{-2}}.
\end{equation}
The masses and widths of the massive particles read \cite{Tanabashi:2018oca}
\begin{alignat}{2}
\label{eqn:ParticleMassesAndWidths}
                  \Mt   &=  173.0\GeV,       & \quad \quad \quad \Gt &= 0 \GeV,  \nonumber \\
                \MZOS &=  91.1876\GeV,      & \quad \quad \quad \GZOS &= 2.4952\GeV,  \nonumber \\
                \MWOS &=  80.379\GeV,       & \GWOS &= 2.085\GeV,  \nonumber \\
                M_{\rm H} &=  125.0\GeV,       &  \GH   &=  4.07 \times 10^{-3}\GeV.
\end{alignat}
The bottom quark is considered massless and is neglected in the initial state by default.
The width of the top quark is set to zero as it is never resonant,
except for the $\mathcal{O}{\left( \alpha^6 \right)}$ contributions
with external bottom quarks, which we consider separately; there we
set the top-quark width to $\Gamma_{\Pt}^\text{LO} = 1.449582\GeV$ \cite{Basso:2015gca}.
The Higgs-boson mass is taken according to the recommendation of the Higgs cross section working group \cite{Heinemeyer:2013tqa} with its corresponding width.
The pole masses and widths entering the calculation are determined
from the measured on-shell (OS) values \cite{Bardin:1988xt} for the W and Z~bosons according to
\begin{equation}
        M_V = \frac{\MVOS}{\sqrt{1+(\GVOS/\MVOS)^2}}\,,\qquad
\Gamma_V = \frac{\GVOS}{\sqrt{1+(\GVOS/\MVOS)^2}}.
\end{equation}

The set of acceptance cuts is taken from the recent CMS measurement,
more precisely, the ones of the \emph{loose fiducial} region defined in \citere{Sirunyan:2019ksz}.
Experimentally, the final state of the process is required to have three charged leptons and at least two jets.
QCD partons are clustered into jets using the anti-$k_\text{T}$
algorithm \cite{Cacciari:2008gp} with jet-resolution parameter
$R=0.4$.  Similarly, photons from real radiation are recombined with
the final-state quarks into jets or with the charged leptons into
dressed leptons, in both cases via the anti-$k_\text{T}$ algorithm and
a resolution parameter $R=0.4$.

In {\sc MoCaNLO} only partons with rapidity $|y|<5$
are considered for recombination, while particles with larger $|y|$ are
assumed to be lost in the beam pipe.
In {\sc BONSAY} all partons are considered for recombination, regardless of their rapidities.
This difference turns out to be numerically irrelevant in our set up.

The pseudo-rapidity $\eta$ and the
transverse momentum $p_{\rm T}$ of a particle are defined as
\begin{align}
\eta = \frac12 \ln \left( \frac{|\vec{p}| + p_z}{|\vec{p}| - p_z} \right) , \qquad
p_{\rm T} = \sqrt{p_x^2+p_y^2},
\end{align}
where $|\vec{p}|$ is the absolute value of the three-momentum~$\vec{p}$
of the particle,
$p_z$ the component of its
momentum along the beam axis, and $p_x$, $p_y$ the components
perpendicular to the beam axis.

The charged leptons $\ell$ are required to pass
the acceptance cuts
\begin{align}
 \ptsub{\Pl} >  20\GeV,\qquad |\eta_{\Pl}| < 2.5, \qquad M_{3\Pl} > 100 \GeV, \qquad M_{\Pl\Pl} > 4 \GeV.
\end{align}
In addition, an invariant-mass cut on the decay products of the Z boson is applied:
\begin{align}
\label{eq:mz}
 | M_{\mu_+ \mu_-} - \MZ | <  15\GeV.
\end{align}
A recombined QCD parton system is called a jet if
it obeys the jet-identification criteria
\begin{align}
 \ptsub{\Pj} >  30\GeV, \qquad |\eta_\Pj| < 4.7,\qquad\Delta R_{\Pj\Pl} > 0.4,
\end{align}
where the last condition requires a minimal distance between a jet and
each of the charged leptons. The identified jets are then ordered according
to the magnitude of their transverse momenta $\ptsub{\Pj,i}$,
where $\ptsub{\Pj,1}$ denotes the largest $\ptsub{\Pj}$ value in the event
and $\ptsub{\Pj,2}$ the second largest.
The distance $\Delta R_{ij}$ between two particles $i$ and $j$ in the pseudo-rapidity--azimuthal-angle plane reads
\begin{equation}
        \Delta R_{ij} = \sqrt{(\Delta \phi_{ij})^2+(\Delta \eta_{ij})^2},
\end{equation}
with $\Delta \phi_{ij}=\min(|\phi_i-\phi_j|,2\pi-|\phi_i-\phi_j|)$
being the azimuthal-angle difference and $\Delta \eta_{ij} = \eta_i-\eta_j$ the
rapidity difference.
On the invariant-mass and rapidity separation of the
leading and sub-leading jets, \emph{i.e.}\ on the two jets with largest
transverse momenta, the following VBS cuts are applied:
\begin{align}
\label{eq:vbscuts}
 M_{\Pj \Pj} >  500\GeV,\qquad |\Delta y_{\Pj \Pj}| > 2.5.
\end{align}

\subsection{Cross sections}

We start our discussion of numerical results by reporting LO cross
sections in the fiducial region.  In \refta{tab:LO} the cross sections
at the orders $\mathcal{O}{\left( \alpha^6 \right)}$,
$\mathcal{O}{\left( \alphas \alpha^5 \right)}$, and
$\mathcal{O}{\left( \alphas^2 \alpha^4 \right)}$ are shown for the
central scale.
\begin{table}
\centering
\begin{tabular}{ccccc}
\toprule
Order
    & $\mathcal{O}{\left(           \alpha^6 \right)}$
    & $\mathcal{O}{\left( \alphas   \alpha^5 \right)}$
    & $\mathcal{O}{\left( \alphas^2 \alpha^4 \right)}$
    & Sum \\
\midrule
$\sigma_{\mathrm{LO}} [\si{\femto\barn}]$
    & \num{0.25511 +- 0.00001}
    & \num{0.006824 +- 0.000001}
    & \num{1.0973 +- 0.0001}
    & \num{1.3592 +- 0.0001} \\
$\Delta [\si{\percent}]$
    & \num{18.8}
    & \num{0.5}
    & \num{80.7}
    & \num{100} \\
\bottomrule
\end{tabular}
\caption{LO cross sections $\sigma_{\rm LO}$ (sum) and individual
orders $\mathcal{O}{\left( \alpha^6 \right)}$,
$\mathcal{O}{\left( \alphas \alpha^5 \right)}$, and
$\mathcal{O}{\left( \alphas^2 \alpha^4 \right)}$
for $\Pp\Pp \to \mu^+\mu^-\Pe^+\nu_\Pe\Pj\Pj+X$ at the LHC with CM energy $13\TeV$.
Photon-induced contributions and contributions with external bottom
quarks are not included.
Each contribution is given in $\fb$ and as fraction $\Delta$
relative to the sum of the three contributions (in percent).
The digits in parentheses indicate the integration errors.}
\label{tab:LO}
\end{table}
In contrast to the like-sign $\PW\PW$ channel, where the EW contribution largely dominates over the QCD one, here the EW contribution is smaller than the QCD contribution by about a factor four.
The LO interference contribution of $\mathcal{O}{\left( \alphas \alpha^5 \right)}$
only amounts to
\SI{0.5}{\percent}
and is, thus, phenomenologically unimportant.

Taking into account scale variation as defined after
\refeq{eq:defscale}, the LO cross section for the quark-induced EW
contribution (often referred to as signal in experimental analysis) is:
\begin{equation}
\sigma_{\rm LO}^{\mathcal{O}{\left( \alpha^6 \right)}} =
\num{0.25511 +- 0.00001}^{+\SI{9.0}{\percent}}_{\SI{-7.8}{\percent}}~\si{\femto\barn}
.
\end{equation}
Note that this order does not involve any strong coupling,
which explains the relatively low scale dependence.
We do not show the scale dependence of the LO contribution at the
orders $\mathcal{O}{\left( \alphas\alpha^5 \right)}$ and $\mathcal{O} (\alpha_\mathrm{s}^2 \alpha^4)$, since the 
corresponding NLO contributions
balancing their scale dependence are not part of this calculation.

In addition to these quark-induced EW contributions, we have also
computed all LO contributions featuring a photon in the initial state.
This includes contributions with initial states $\Pg\gamma$,
$\Pq\gamma$ as well as $\gamma\gamma$ at orders 
$\mathcal{O}{\left( \alpha^6 \right)}$ or $\mathcal{O}{\left( \alphas \alpha^5
  \right)}$.  As can be seen from \refta{tab:LOsub}, these
contributions are phenomenologically negligible.  In addition, the LO
contributions at order $\mathcal{O}{\left( \alpha^6 \right)}$
involving bottom quarks either in the initial state or in the final
state are also reported.  While the contributions with two bottom
quarks in the initial state are negligible due to their PDF
suppression, the contributions with one light quark and one bottom
quark in the initial state are rather large.  The latter are usually
referred to as $\Pt\PZ+\mathrm{jet}$ contributions in experimental
analyses (see \reffi{fig:born_bq_tZj}).  These contributions are
enhanced due to resonant top-quark contributions.  In the final state
they have one b-jet and one light jet and can therefore be suppressed
in experimental analyses 
using b-jet tagging techniques.\footnote{In the $\PW\PZ$ analysis of \citere{Sirunyan:2019ksz}, such $\Pt\PZ+\mathrm{jet}$ contributions are suppressed by a central b-jet veto for $|\eta|<2.5$.
The residual contribution is then estimated from Monte Carlo simulations and subtracted as background.
Conversely, the $\Pt\PZ+\mathrm{jet}$ process has recently been observed in \citere{Sirunyan:2018zgs} where the $\PW\PZ$ EW contribution is considered as background.
We have verified by a LO calculation that $91\%$ of the tZ+jet contribution has a leading b-jet contained within $|\eta_b|<2.5$.}
Note that
these contributions also contain VBS contributions
(for instance diagram \reffi{fig:born_qq_vbs} with the lower up-quark
line replaced by a bottom quark line), but are dominated by
contributions of a resonant top quark.

\begin{table}
\centering
\begin{tabular}{ccc}
\toprule
Contribution
    & $\gamma$-induced
    & bottom \\
\midrule
$\Delta\sigma_{\mathrm{LO}} [\si{\femto\barn}]$
    & \num{0.0009884 +- 0.0000002}
    & \num{0.19451 +- 0.00002} \\
$\Delta\sigma_{\mathrm{LO}} / \sigma_{\mathrm{LO}}^{\mathcal{O}{\left( \alpha^6 \right)}} [\si{\percent}]$
    & \num{0.4}
    & \num{76.2} \\
\bottomrule
\end{tabular}
\caption{LO cross-section contributions $\Delta\sigma_{\rm LO}$ for $\Pp\Pp \to \mu^+\mu^-\Pe^+\nu_\Pe\Pj\Pj+X$
with initial-state photons or external bottom quarks.
The photon-induced contributions involve one or two initial-state photons
and contribute to the orders $\mathcal{O}{\left( \alpha^6 \right)}$ and
$\mathcal{O}{\left( \alphas \alpha^5 \right)}$.
The ``bottom'' contributions are of the order
$\mathcal{O}{\left( \alpha^6 \right)}$ and $\mathcal{O} (\alpha_\mathrm{s}^2 \alpha^4)$, and involve bottom quarks
in the initial and/or final state.
All contributions are given in $\fb$ as well as relative
to the LO EW cross section of order $\mathcal{O}{\left( \alpha^6
  \right)}$ (in percent).
The digits in parentheses indicate the integration errors.}
\label{tab:LOsub}
\end{table}

\begin{table}
\sisetup{group-digits=false}
\centering
\begin{tabular}{cccc}
\toprule
Order
    & $\mathcal{O}{\left(         \alpha^6 \right)}+\mathcal{O}{\left(         \alpha^7 \right)}$
    & $\mathcal{O}{\left(         \alpha^6 \right)}+\mathcal{O}{\left( \alphas \alpha^6 \right)}$
    & $\mathcal{O}{\left(         \alpha^6 \right)}+\mathcal{O}{\left(         \alpha^7 \right)} + \mathcal{O}{\left( \alphas \alpha^6 \right)}$ \\
\midrule
$\sigma_{\mathrm{NLO}} [\si{\femto\barn}]$
    & \num{0.2142 +- 0.0002} 
    & \num{0.2506 +- 0.0001} 
    & \num{0.2097 +- 0.0003} \\
$\sigma^{\mathrm{max}}_{\mathrm{NLO}} [\si{\femto\barn}]$
    & \num{0.2325 +- 0.0003} $[+8.5\%]$
    & \num{0.2532 +- 0.0001} $[+1.0\%]$
    & \num{0.2125 +- 0.0002} $[+1.3\%]$\\
$\sigma^{\mathrm{min}}_{\mathrm{NLO}} [\si{\femto\barn}]$
    & \num{0.1984 +- 0.0002} $[-7.4\%]$
    & \num{0.2481 +- 0.0001} $[-1.0\%]$
    & \num{0.2050 +- 0.0003} $[-2.2\%]$\\
$\delta [\si{\percent}]$
    & \num{-16.0}
    & \num{-1.8}
    & \num{-17.8} \\
\bottomrule
\end{tabular}
\caption{
Cross sections for $\Pp\Pp \to \mu^+\mu^-\Pe^+\nu_\Pe\Pj\Pj+X$ at the
LHC with CM energy $13\TeV$ at NLO EW  [$\mathcal{O}{\left(\alpha^6
  \right)}+\mathcal{O}{\left(\alpha^7 \right)}$], 
  NLO QCD [$\mathcal{O}{\left(\alpha^6\right)}
  + \mathcal{O}{\left(\alphas \alpha^6 \right)}$], 
  and NLO QCD+EW  [$\mathcal{O}{\left(\alpha^6 \right)}
  +\mathcal{O}{\left(\alpha^7 \right)} + \mathcal{O}{\left( \alphas \alpha^6 \right)}$].
Each contribution is given in $\fb$ (with the extrema resulting from
scale variations as absolute numbers and as deviation in percent) and
as relative correction 
$\delta=\sigma_{\rm NLO} / \sigma_{\rm LO}^{\mathcal{O}{\left( \alpha^6 \right)}}-1$
to the LO EW cross section
of order $\mathcal{O}{\left( \alpha^6 \right)}$ in percent.
The digits in parentheses indicate the integration errors.}
\label{tab:NLO}
\end{table}

NLO cross sections including orders $\mathcal{O}{\left(\alphas\alpha^6\right)}$ or/and $\mathcal{O}{\left( \alpha^7 \right)}$ in addition to the LO $\mathcal{O}{\left( \alpha^6 \right)}$ are reported in
\refta{tab:NLO} for the central scale as well as with the two extrema
resulting from the 7-point scale variation.  If only the
$\mathcal{O}{\left( \alpha^7 \right)}$ corrections are included the
scale uncertainty remains at the same level as in LO, while the
inclusion of the $\mathcal{O}{\left( \alphas \alpha^6 \right)}$
corrections reduces the scale uncertainty as expected.
The NLO contribution of
order $\mathcal{O}{\left( \alphas \alpha^6 \right)}$ amounts to about
\SI{-1.8}{\percent}
with respect to the LO of order $\mathcal{O}{\left( \alpha^6
  \right)}$.  As explained previously, this correction is of mixed
type, \emph{i.e.}\ it features both QCD and EW corrections.
Nonetheless it is often referred to as QCD correction to the EW
signal, as the VBS approximation neglects the (comparably small) EW
corrections of order $\mathcal{O}{\left(\alphas\alpha^6\right)}$.

On the other hand, the EW corrections of order $\mathcal{O}{\left(
  \alpha^7 \right)}$ amount to
\SI{-16}{\percent}
and represent the dominant NLO contribution.  This is in line with the
findings of \citere{Biedermann:2016yds} for like-sign WW~scattering
and supports the expectation that large EW corrections are an
intrinsic feature of VBS at the LHC.  Following
\citere{Denner:2000jv}, one can derive a leading logarithmic
approximation for the EW corrections to the process $\Pp \Pp \to
\mu^+\mu^-\Pe^+\nu_\Pe\Pj\Pj+X$ 
based on the logarithmic corrections to
the sub-process $\PW\PZ \to \PW\PZ$.
Taking the mixing of photon and Z~boson into
account and using $\mathcal{M}^{\PW\PZ\to\PW\gamma} \approx -
\frac{\sw}{\cw} \mathcal{M}^{\PW\PZ\to\PW\PZ}$, one arrives at the
approximation already given in \citere{Biedermann:2016yds} for
$\PW\PW\to\PW\PW$. This approximation holds in fact for all scattering
processes of EW bosons owing to the fact that these scattering
processes result from the same $\mathrm{SU}(2)_\mathrm{w}$ coupling. The
approximation reads
\begin{equation}
\rd\sigma_{\textrm{LL}} = \rd\sigma_{\textrm{LO}}
\left( 1 + \delta_{\textrm{EW,LL}} \right) ,
\end{equation}
where
\begin{equation}
 \delta_{\textrm{EW,LL}} = \frac{\alpha}{4\pi} \left\{
- 4 C_W^{\mathrm{EW}} \log^2 \left(\frac{Q^2}{\MW^2}\right)
+  2b_W^{\mathrm{EW}} \log \left(\frac{Q^2}{\MW^2}\right) \right\}
\label{eq:LLcorr}
\end{equation}
with $ C_W^{\mathrm{EW}} = 2/\sw^2$ and $ b_W^{\mathrm{EW}} =
19/(6\sw^2)$.  The symbols $\cw$ and $\sw$ represent the cosine and
sine of the weak mixing angle, respectively.  The scale $Q$ is a
representative scale of the $\PW\PZ \to \PW\PZ$ scattering process;
the four-lepton invariant mass $M_{4\ell}$ turns out to be
particularly appropriate.  Setting $Q$ to the average LO value
$\langle M_{4\ell}\rangle \simeq 413 \GeV$ and applying
\refeq{eq:LLcorr} to the integrated cross section, leads to a leading
logarithmic correction of $\delta_{\textrm{EW,LL}} = -17.5 \%$, which
is good given the approximation used.  Applying $Q=M_{4\ell}$ event by
event in the calculation results in $\delta_{\textrm{EW,LL}} = -16.4
\%$, which agrees even better with the result of the full calculation.
As already noted in \citere{Biedermann:2016yds}, the
  rather large average scale $\langle M_{4\ell}\rangle$ for VBS
  processes is not due to the peculiar VBS event selection but to an
  enhancement of the partonic $qq'$ cross section containing the
  $VV' \to VV'$ subprocess resulting from a massive $t$-channel exchange
  \cite{Denner:1997kq}. It was verified for the related $\PW^+\PW^+$
  scattering process that relaxing the cuts leaves the EW corrections
  at the same level.

Finally, the fiducial cross section with both NLO QCD and EW corrections added is
\begin{equation}
\sigma_{\rm NLO}^{\rm{QCD}+\rm{EW}} =
\num{0.2097 +- 0.0003}^{+\SI{1.3}{\percent}}_{\SI{-2.2}{\percent}}~\si{\femto\barn}
,
\end{equation}
showing a significant reduction of scale uncertainty.
This is mainly due to the $\mathcal{O}(\alphas)$ PDF redefinition included in
the $\mathcal{O}(\alphas \alpha^6)$ NLO correction that cancels the factorisation scale dependence
of the LO $\mathcal{O}(\alpha^6)$ contribution.
As shown in
\refta{tab:NLO}, the full NLO correction is about
\SI{-17.8}{\percent}
with respect to the LO of order $\mathcal{O}{\left( \alpha^6 \right)}$.

Finally, for completeness, we also provide cross sections at NLO for the LHC running at a centre-of-mass energy of $14\TeV$ in Table~\ref{tab:NLO14}.
While the LO cross section increases by $17.2\%$ with respect to $13\TeV$, the relative NLO corrections are rather stable.
These numbers can be important for future operation of the LHC at high luminosity \cite{Azzi:2019yne} and serve as benchmarks.

\begin{table}
\centering
\begin{tabular}
{ccccc}
\toprule
 Order &  $\mathcal{O}{\left( \alpha^6 \right)}$ &  $\mathcal{O}{\left(\alpha^6 \right)}+\mathcal{O}{\left( \alpha^7 \right)}$ & 
$\mathcal{O}{\left(\alpha^6 \right)}+\mathcal{O}{\left( \alphas
    \alpha^6 \right)}$ & NLO QCD+EW 
\\
\midrule
$\sigma [\fb]$           & $0.2988(6)$ & $0.251(1)$ &  $0.294(1)$ & $0.245(2)$ \\
$\sigma^{\rm max} [\fb]$ & $0.3244(6) [+8.5\%]$ & $0.271(1) [+8.0\%]$ & $0.296(1) [+0.7\%]$ & $0.247(1) [+0.8\%]$ \\
$\sigma^{\rm min} [\fb]$ & $0.2767(6) [-7.4\%]$ & $0.233(1) [-7.2\%]$ & $0.291(1) [-1.0\%]$ & $0.243(2) [-0.8\%]$ \\
$\delta [\%]$ & --- & $-16.1$ & $-1.8$ & $-17.9$ \\
\bottomrule
\end{tabular}
\caption{
Cross sections for $\Pp\Pp \to \mu^+\mu^-\Pe^+\nu_\Pe\Pj\Pj+X$ at the
LHC with CM energy $14\TeV$ at LO [$\mathcal{O}{\left( \alpha^6
  \right)}$], 
  NLO EW  [$\mathcal{O}{\left(\alpha^6
  \right)}+\mathcal{O}{\left(\alpha^7 \right)}$], 
  NLO QCD [$\mathcal{O}{\left(\alpha^6\right)}
  + \mathcal{O}{\left(\alphas \alpha^6 \right)}$], 
  and NLO QCD+EW  [$\mathcal{O}{\left(\alpha^6 \right)}
  +\mathcal{O}{\left(\alpha^7 \right)} + \mathcal{O}{\left( \alphas \alpha^6 \right)}$].
Each contribution is given in $\fb$ (with the extrema resulting from
scale variations as absolute numbers and as deviation in percent) and
as relative correction 
$\delta=\sigma_{\rm NLO} / \sigma_{\rm LO}^{\mathcal{O}{\left( \alpha^6 \right)}}-1$
to the LO EW cross section
of order $\mathcal{O}{\left( \alpha^6 \right)}$ in percent.
The digits in parentheses indicate the integration errors.}
\label{tab:NLO14}
\end{table}

\subsection{Differential distributions}

\begin{figure}
        \setlength{\parskip}{-10pt}
        \begin{subfigure}{0.49\textwidth}
                \subcaption{}
                 \includegraphics[width=\textwidth,page=14]{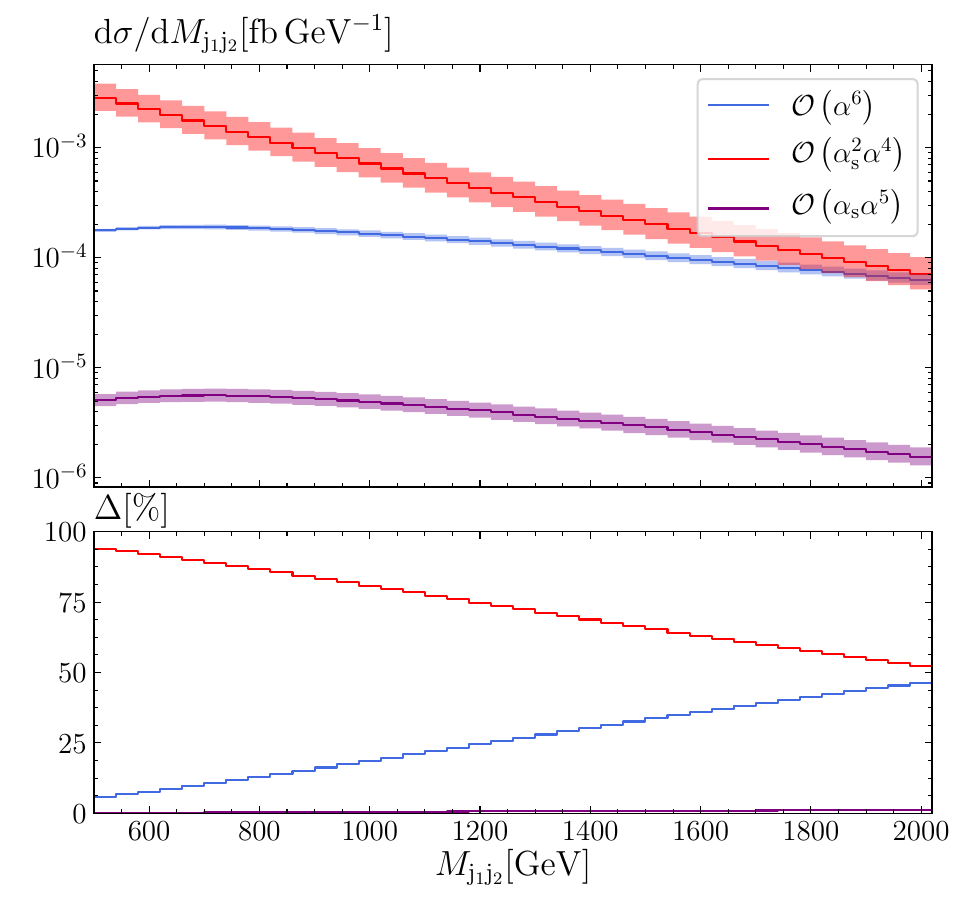}
                \label{plot:mjj_lo}
        \end{subfigure}
        \hfill
        \begin{subfigure}{0.49\textwidth}
                \subcaption{}
                 \includegraphics[width=\textwidth,page=2]{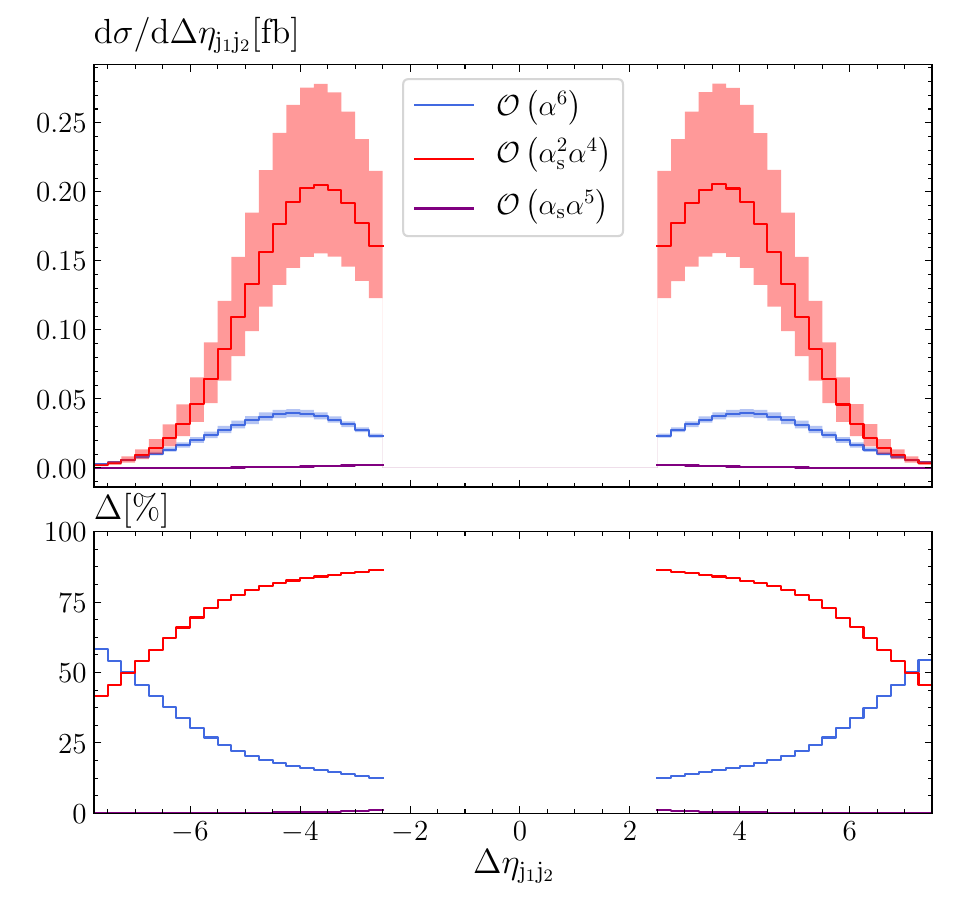}
                \label{plot:dyjj_lo}
        \end{subfigure}

        \begin{subfigure}{0.49\textwidth}
                \subcaption{}
                 \includegraphics[width=\textwidth,page=21]{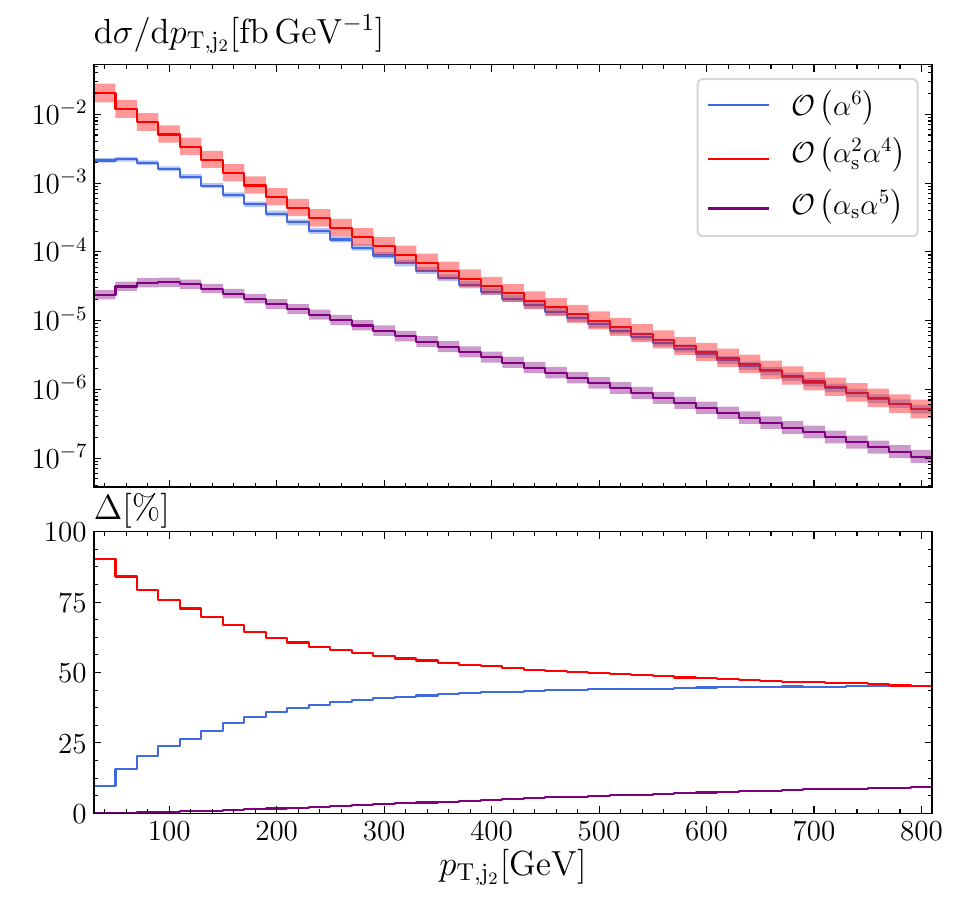}
                \label{plot:pT_j2_lo}
        \end{subfigure}
        \hfill
        \begin{subfigure}{0.49\textwidth}
                \subcaption{}
                 \includegraphics[width=\textwidth,page=5]{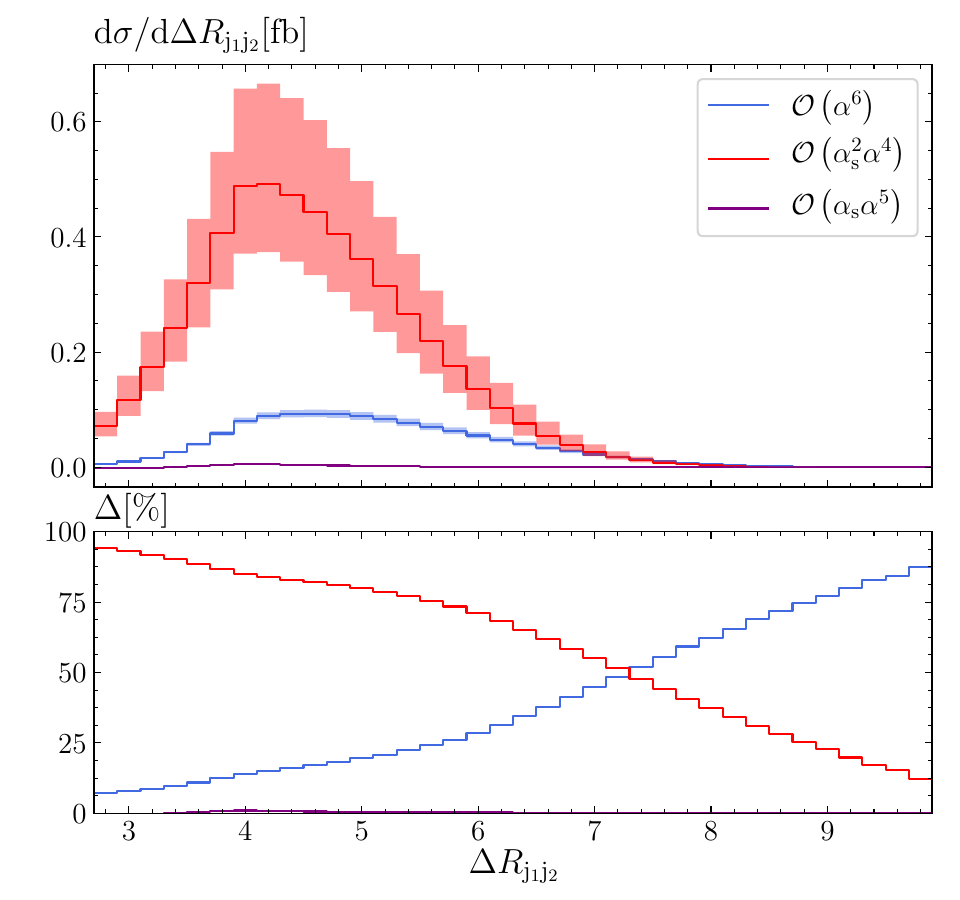}
                \label{plot:drjj_lo}
        \end{subfigure}

        \vspace*{-3ex}
        \caption{\label{fig:dist_lo}%
                LO differential distributions for $\Pp\Pp \to \mu^+\mu^-\Pe^+\nu_\Pe\Pj\Pj+X$ 
                at the LHC with CM energy $13\TeV$:
                \subref{plot:mjj_lo}~invariant mass of the two jets~(top left), %
                \subref{plot:dyjj_lo}~difference of pseudo-rapidity of the two jets~(top right), 
                \subref{plot:pT_j2_lo}~transverse momentum of the second hardest jet~(bottom left), and
                \subref{plot:drjj_lo}~rapidity--azimuthal-angle distance between the two jets~(bottom right).
                The upper panel shows the absolute contributions of order $\mathcal{O}{\left( \alpha^6 \right)}$ (EW), 
                $\mathcal{O}{\left( \alphas \alpha^5 \right)}$ (interference), and $\mathcal{O}{\left( \alphas^2 \alpha^4 \right)}$ (QCD).
                The bands denote the envelope of the scale variation for each order.
                The lower panel shows the relative LO contributions $\Delta$ to their sum in percent.}
\end{figure}

In this section, LO predictions and NLO corrections for several
differential distributions are discussed.  We start with 
a few LO predictions in \reffi{fig:dist_lo}.  The upper panels show the
absolute predictions of order $\mathcal{O}{\left( \alpha^6 \right)}$
(EW), $\mathcal{O}{\left( \alphas \alpha^5 \right)}$ (interference),
and $\mathcal{O}{\left( \alphas^2 \alpha^4 \right)}$ (QCD).  In the
lower panels, the relative contributions are displayed with respect to
the sum of the three contributions.  Note that the contributions
featuring external bottom quarks or initial-state photons are not included here.
The first two distributions are the invariant mass and pseudo-rapidity
difference of the two tagging jets in \reffis{plot:mjj_lo} and
\ref{plot:dyjj_lo}.  These observables are often used to separate EW
and QCD contributions in experimental analysis.  This is perfectly
justified by the fact that at higher invariant mass or larger
pseudo-rapidity, the EW contribution is becoming dominant.  The effect
of the event selection for $|\Delta \eta_{\Pj_1\Pj_2}| > 2.5$ and
$|M_{\Pj_1\Pj_2}| > 500\GeV$ is clearly visible.  In
\reffi{plot:pT_j2_lo}, the transverse momentum of the second hardest
jet is shown.  Around $500\GeV$, both the EW and QCD contributions
become of the same size, as the QCD contribution is falling much more
steeply than the EW contribution.  Interestingly, this behaviour is
not visible in other transverse-momentum and invariant-mass
distributions, like the transverse-momentum distribution of the
leading jet, where the QCD contributions are always larger than the EW
contributions.  The comparably steep fall of the distribution in the
transverse momentum of the subleading jet is due to the fact that QCD
contributions are dominated by contributions with at least one jet
with small transverse momentum. Finally, \reffi{plot:drjj_lo} displays the
distribution in the rapidity--azimuthal-angle distance between the
leading jets which also shows a good discriminating power, as already 
noticed in \citere{Bendavid:2018nar}.  Note that in all
distributions, the interference contribution is very much suppressed
reflecting its overall small cross section.

The following figures show our results on NLO differential distributions. 
In the upper panels, the LO contribution of order $\mathcal{O}{\left( \alpha^6 \right)}$ is shown
along with the NLO
predictions including orders $\mathcal{O}{\left(\alpha^7\right)}$ or/and $\mathcal{O}{\left( \alphas \alpha^6 \right)}$.
For simplicity, these are often denoted by EW and QCD corrections,
respectively, in the following.  We stress again that while the order
$\mathcal{O}{\left(\alpha^7 \right)}$ comprises genuine EW
corrections, the order $\mathcal{O}{\left( \alphas \alpha^6 \right)}$
contains both QCD and EW corrections.  We adopt this assignment in
order to facilitate the reading. For the QCD corrections, the lower panels show the relative
contributions
\begin{equation}
\delta = \frac{\rd\sigma(\mu)}
{\rd\sigma_{\rm LO}^{\mathcal{O}{\left( \alpha^6 \right)}}(\mu_0)}-1,
\end{equation}
where the bands in the plots reflect the variation of the numerator with the
(renormalisation and/or factorisation) scale $\mu$ while keeping the
scales in the denominator fixed to $\mu_0$.
For the EW corrections, only the value for the central scale is shown
in the lower panels, because the scale dependence of the corresponding
relative NLO contribution is negligible against the one of the other
contributions. The larger scale variation of NLO QCD+EW with respect to 
NLO QCD in the plots results from the inclusion of the large EW NLO correction 
in the numerator of (4.15), where the $\mu$-insensitive relative EW correction 
multiplies the scale-dependent LO cross section.

\begin{figure}
        \setlength{\parskip}{-10pt}
        \begin{subfigure}{0.49\textwidth}
                \subcaption{}
                 \includegraphics[width=\textwidth,page=19]{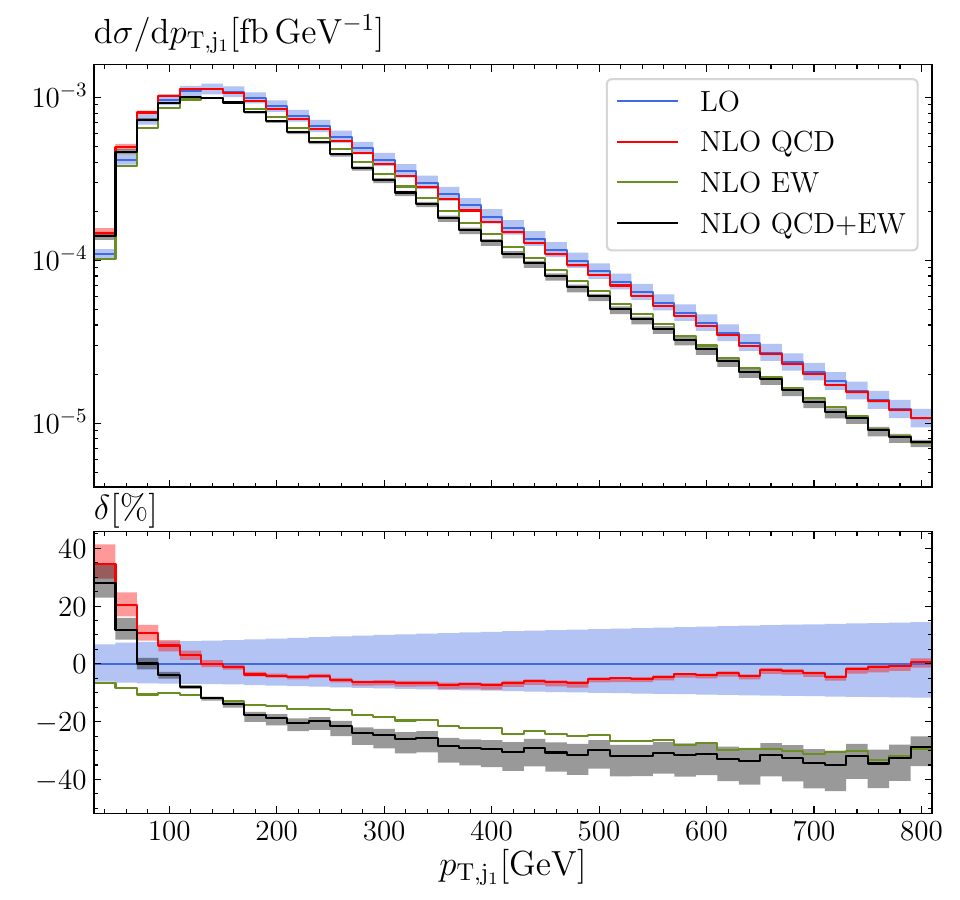}
                \label{plot:pT_j1}
        \end{subfigure}
        \hfill
        \begin{subfigure}{0.49\textwidth}
                \subcaption{}
                 \includegraphics[width=\textwidth,page=21]{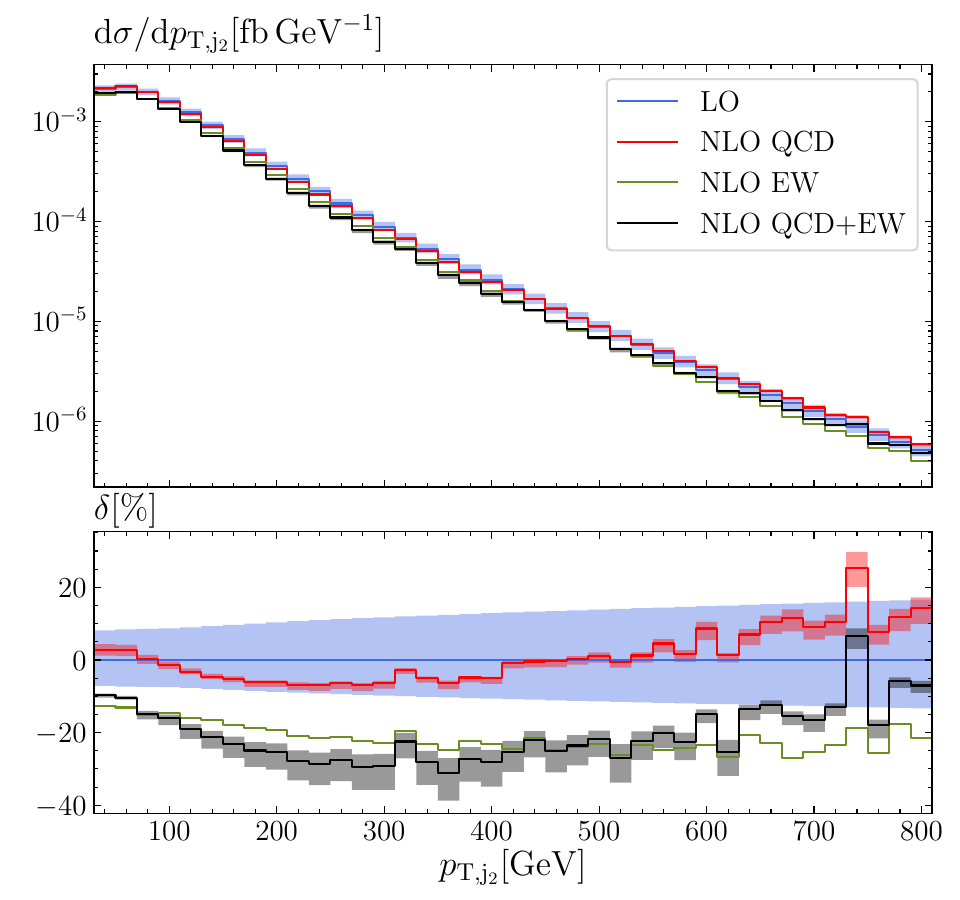}
                \label{plot:pT_j2}
        \end{subfigure}

        \begin{subfigure}{0.49\textwidth}
                \subcaption{}
                 \includegraphics[width=\textwidth,page=22]{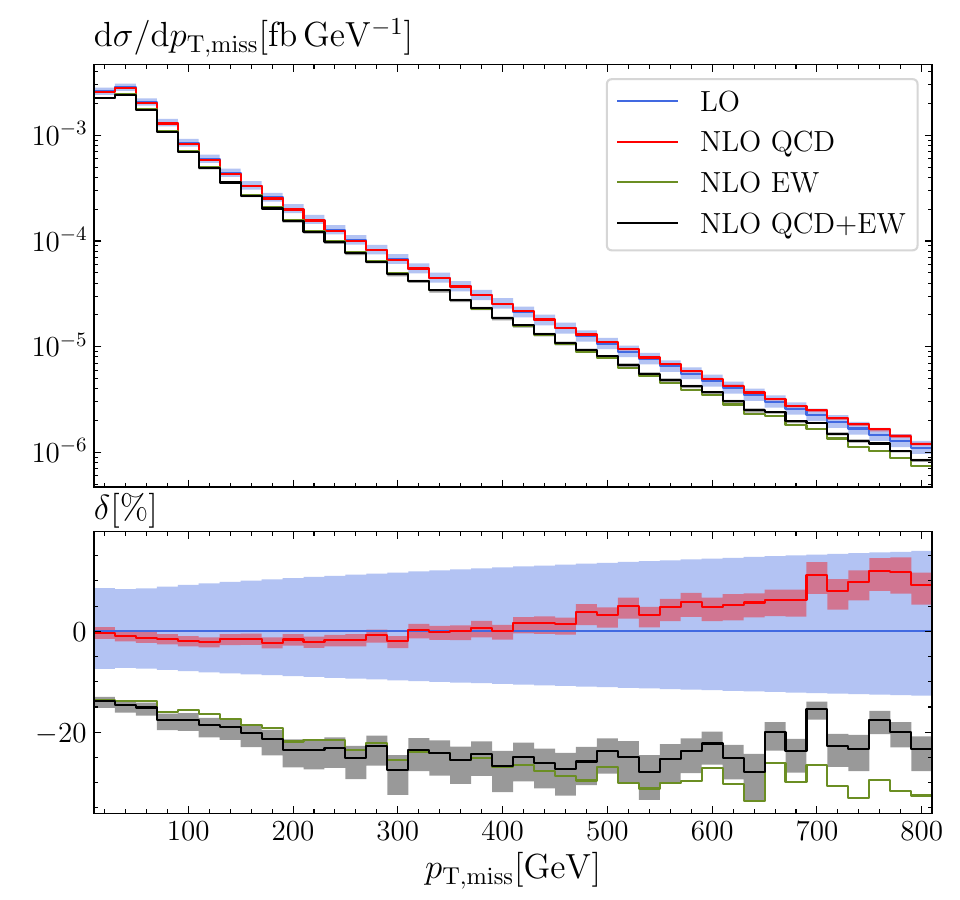}
                \label{plot:pT_miss}
        \end{subfigure}
        \hfill
        \begin{subfigure}{0.49\textwidth}
                \subcaption{}
                 \includegraphics[width=\textwidth,page=18]{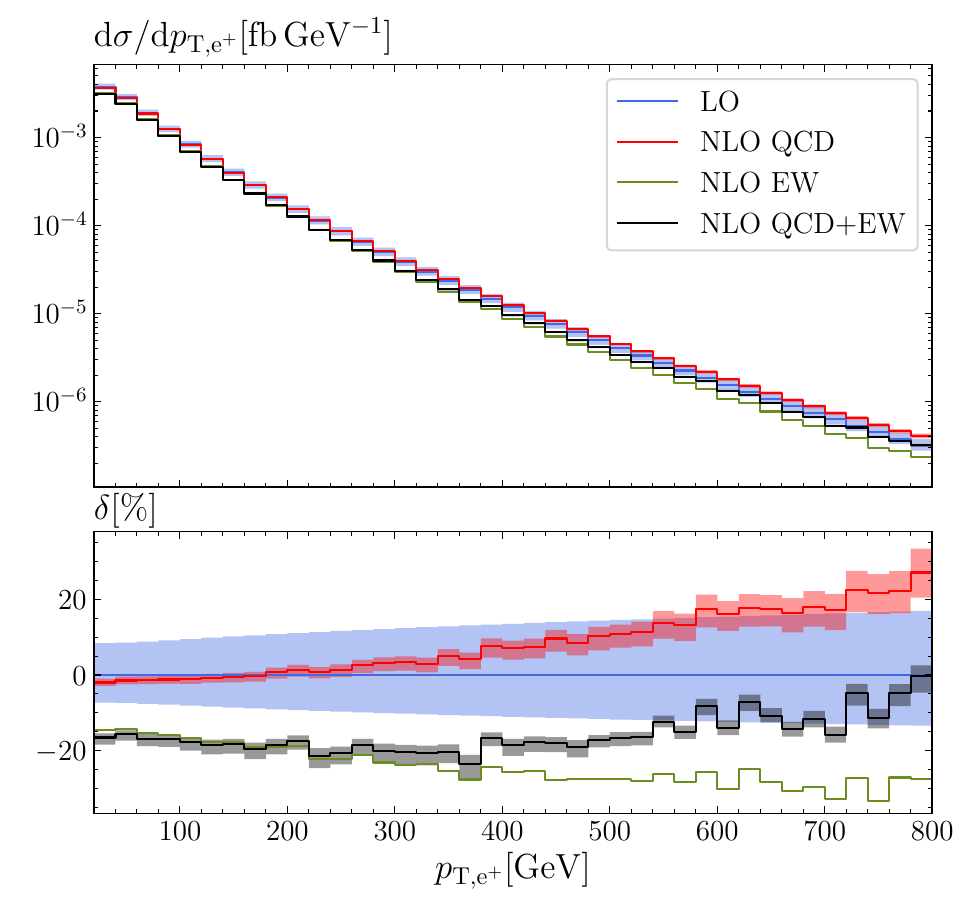}
                \label{plot:pT_e}
        \end{subfigure}

        \vspace*{-3ex}
        \caption{\label{fig:dist_pt}%
                Differential distributions for $\Pp\Pp \to \mu^+\mu^-\Pe^+\nu_\Pe\Pj\Pj+X$ at the LHC with CM energy $13\TeV$:
                \subref{plot:pT_j1}~transverse momentum of the hardest jet~(top left), %
                \subref{plot:pT_j2}~transverse momentum of the second
                hardest jet~(top right), \subref{plot:pT_miss}~missing
                transverse energy~(bottom left), and
                \subref{plot:pT_e}~transverse momentum of the
                positron~(bottom right).  The upper panel shows the LO
                contributions of order $\mathcal{O}{\left( \alpha^6
                  \right)}$, the two NLO predictions
                [including $\mathcal{O}{\left(\alpha^7 \right)}$ (NLO EW) and
                  $\mathcal{O}{\left( \alphas \alpha^6 \right)}$ (NLO
                  QCD)] as well as their sum.  The lower panel shows
                  the relative NLO corrections with respect to the LO
                  in percent.}
\end{figure}
In \reffi{fig:dist_pt}, several distributions in transverse momenta
are presented.  We start with those for the hardest and second hardest
jet in \reffis{plot:pT_j1} and \ref{plot:pT_j2}, respectively.
For both distributions, the EW corrections become large
  in size and negative for large transverse momenta.  The QCD
corrections are positive for low transverse momentum of the leading
jet, but steadily decrease towards high transverse momentum, becoming
negative above $150\GeV$ and stabilising in the range $200{-}300\GeV$.
This behaviour is typical of a process with hard jet
  emission in its signature and results from the reduction of the
  leading-jet transverse momentum by emission of real gluons and has
  also been observed in like-sign WW
  scattering~\cite{Biedermann:2017bss}.  For the transverse momentum
of the second leading jet, the QCD corrections turn again positive
towards high transverse momentum.  The enhanced corrections for small
transverse momentum of the leading jet are due to the phase-space
suppression of the LO when all jet transverse momenta are required to
be small. This causes corrections above $20\%$ for small transverse
momenta of the hardest jet, while the corrections almost vanish for
small transverse momenta of the second hardest jet.
For the distributions in the missing transverse momentum
(\reffi{plot:pT_miss}), which is identified with the neutrino
transverse momentum $p_{\mathrm{T},\nu_\Pe}$, and in the transverse
momentum of the positron $p_{\mathrm{T},\Pe^+}$ (\reffi{plot:pT_e}),
the EW corrections increase negatively towards higher
  transverse momenta and exceed $-25\%$ at $p_{\rT}=800\GeV$.  The
QCD corrections are almost independent of $p_{\rT,\mathrm{miss}}$ and
$p_{\mathrm{T},\Pe^+}$ until about $400\GeV$.

\begin{figure}
        \setlength{\parskip}{-10pt}
        \begin{subfigure}{0.49\textwidth}
                \subcaption{}
                 \includegraphics[width=\textwidth,page=24]{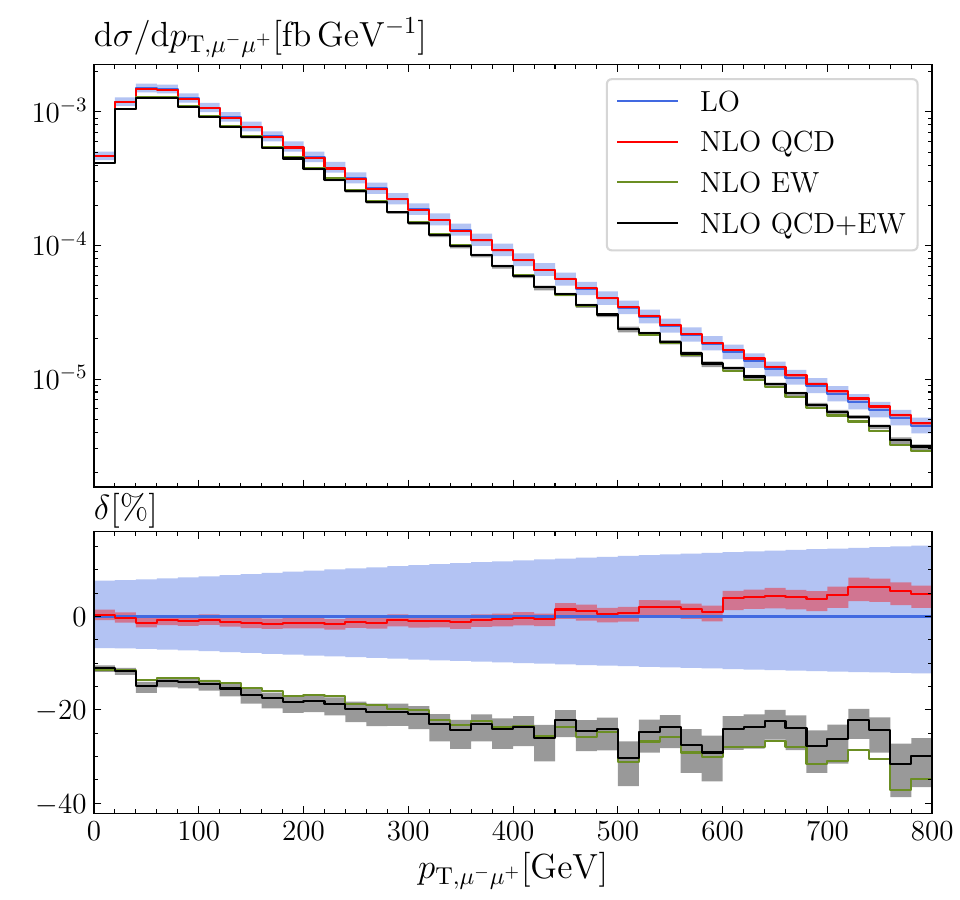}
                \label{plot:pT_Z}
        \end{subfigure}
        \hfill
        \begin{subfigure}{0.49\textwidth}
                \subcaption{}
                 \includegraphics[width=\textwidth,page=26]{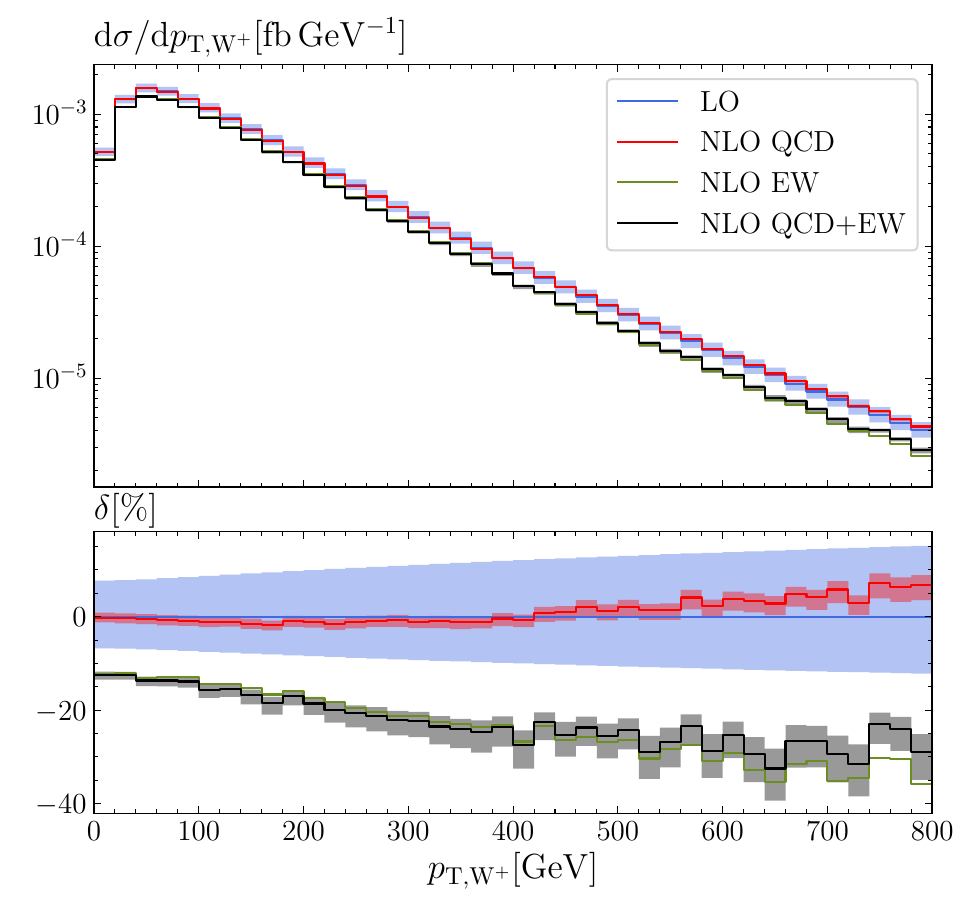}
                \label{plot:pT_W}
        \end{subfigure}
        
        \vspace*{-3ex}
        \caption{\label{fig:dist_pt2}%
                Differential distributions for $\Pp\Pp \to \mu^+\mu^-\Pe^+\nu_\Pe\Pj\Pj+X$ at the LHC with CM energy $13\TeV$:
                \subref{plot:pT_Z}~transverse momentum of the muon--anti-muon system~(left) and%
                \subref{plot:pT_W}~transverse momentum of the reconstructed $\PW$ boson~(right). 
                The upper panel shows the LO contributions of order $\mathcal{O}{\left( \alpha^6
                  \right)}$, the two NLO predictions
                [including $\mathcal{O}{\left(\alpha^7 \right)}$ (NLO EW) and
                  $\mathcal{O}{\left( \alphas \alpha^6 \right)}$ (NLO
                  QCD)] as well as their sum.  The lower panel shows
                  the relative NLO corrections with respect to the LO
                  in percent.}
\end{figure}
Since only the transverse momentum of one final-state
  particle becomes large in the distributions in \reffi{fig:dist_pt},
  the dominant kinematics is not necessarily in the Sudakov region,
  where all invariants are large. On the other hand, if the transverse
  momentum of one of the reconstructed gauge bosons gets large, the
  invariants of the dominating $VV'\to VV'$ scattering subprocess
  become large, and the Sudakov approximation applies to this
  subprocess. The distributions in the transverse momentum of the
  muon--anti-muon system shown in \reffi{plot:pT_Z} and in the
  reconstructed transverse momentum of the $\PW$ boson in
  \reffi{plot:pT_W} indeed display the typical Sudakov behaviour more clearly.
  The EW corrections rise monotonically to $-35\%$ at $p_{\rT}=800\GeV$.
  The QCD corrections, on the other hand, have a limited impact,
  reaching a maximum of only about $5\%$ at $800\GeV$.
 
 In addition to the transverse momentum distributions of
 Fig.~\ref{fig:dist_pt}, we also show the reconstructed transverse
 momentum of the two gauge bosons in Fig.~~\ref{fig:dist_pt2}.  In
 Fig.~\ref{plot:pT_Z}, the transverse momentum of the muon--anti-muon
 system is shown while Fig.~\ref{plot:pT_W} displays the reconstructed
 transverse momentum of the $\PW$ boson.  The distributions display a
 similar behaviour for both the QCD and EW corrections.  The QCD
 corrections have a limited impact, reaching a maximum of only about
 $5\%$ at $800\GeV$.  On the other hand, the EW corrections show a
 monotonic behaviour, increasing negatively to reach $-35\%$ at
 $800\GeV$.  Such a behaviour is typical of EW Sudakov logarithms
 becoming large in the high-energy limit.  Note that the other
 distributions shown later do not show such a pronounced
 Sudakov-logarithmic behaviour as they naturally inherit the intrinsic
 scale of the process.

\begin{figure}
        \setlength{\parskip}{-10pt}
        \begin{subfigure}{0.49\textwidth}
                \subcaption{}
                 \includegraphics[width=\textwidth,page=3]{Figures/nlos}
                \label{plot:dphi_j1j2}
        \end{subfigure}
        \hfill
        \begin{subfigure}{0.49\textwidth}
                \subcaption{}
                 \includegraphics[width=\textwidth,page=5]{Figures/nlos}
                \label{plot:dR_j1j2}
        \end{subfigure}

        \begin{subfigure}{0.49\textwidth}
                \subcaption{}
                 \includegraphics[width=\textwidth,page=6]{Figures/nlos}
                \label{plot:dR_mumu}
        \end{subfigure}
        \hfill
        \begin{subfigure}{0.49\textwidth}
                \subcaption{}
                 \includegraphics[width=\textwidth,page=1]{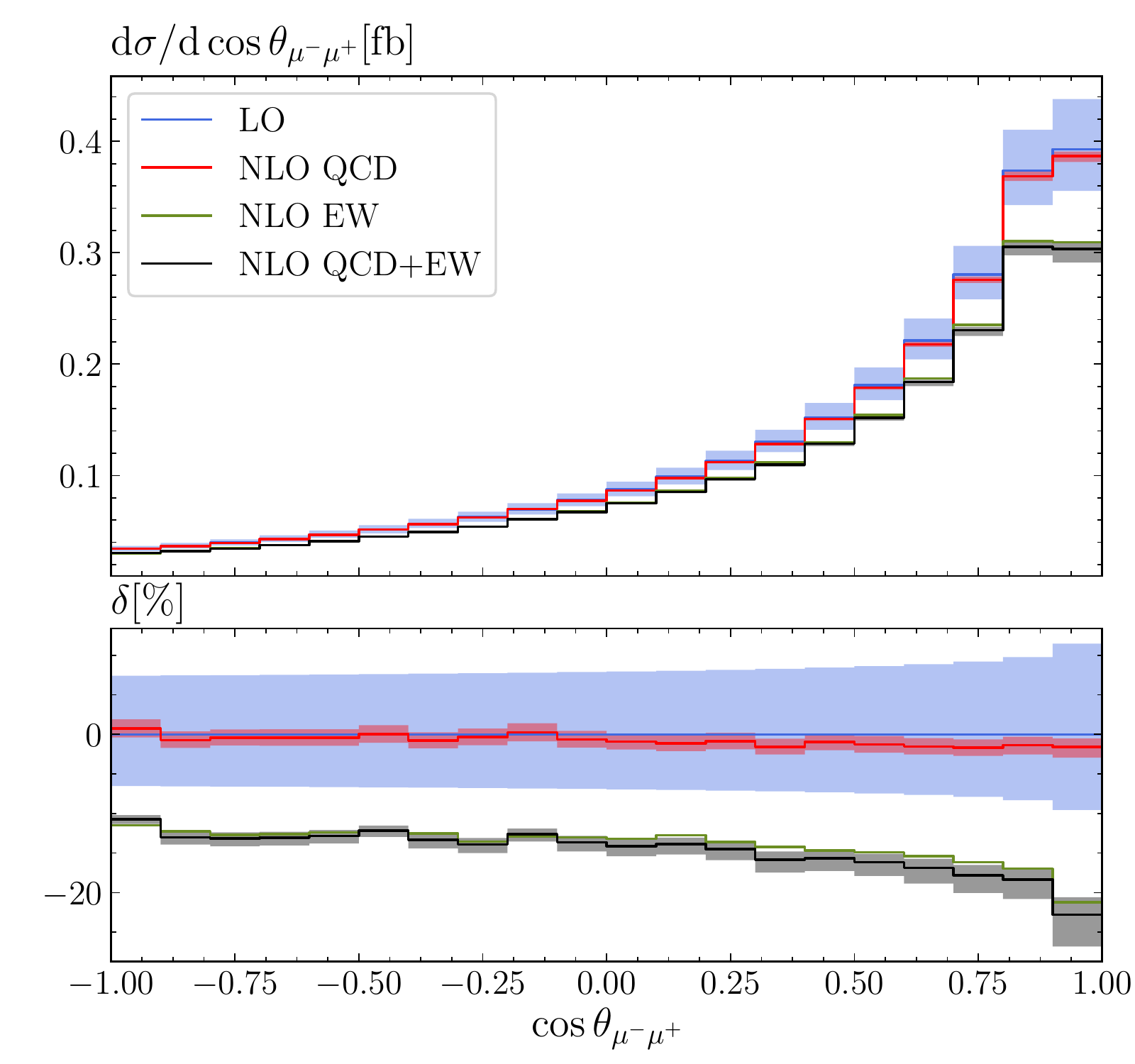}
                \label{plot:cosphi_mumu}
        \end{subfigure}

        \vspace*{-3ex}
        \caption{\label{fig:angle}%
                Differential distributions for $\Pp\Pp \to \mu^+\mu^-\Pe^+\nu_\Pe\Pj\Pj+X$ at the LHC with CM energy $13\TeV$:
                \subref{plot:dphi_j1j2}~azimuthal angle between the
                two tagging jets~(top left),
                \subref{plot:dR_j1j2}~rapidity--azimuthal-angle
                distance between the two tagging jets~(top right), %
                \subref{plot:dR_mumu}~rapidity--azimuthal-angle 
                distance between the muon and
                anti-muon~(bottom left), and
                \subref{plot:cosphi_mumu}~cosine of the angle between
                the muon and anti-muon~(bottom right).  The upper
                panel shows the LO contributions of order
                $\mathcal{O}{\left( \alpha^6 \right)}$, the two NLO
                  predictions [including $\mathcal{O}{\left(\alpha^7 \right)}$
                    (NLO EW) and $\mathcal{O}{\left( \alphas \alpha^6
                      \right)}$ (NLO QCD)] as well as their sum.  The
                    lower panel shows the relative NLO corrections
                    with respect to the LO in percent.}
\end{figure}
Figure~\ref{fig:angle} displays some angular distributions.  For the
distribution in the azimuthal-angle difference of the two tagging jets
(\reffi{plot:dphi_j1j2}) QCD and EW corrections follow a similar trend
and inherit mostly the corrections to the total cross section.  This does
not hold for the other angular distributions, where the corrections
show differences in shape.  For the rapidity--azimuthal-angle distance
between the two jets (\reffi{plot:dR_j1j2}), the QCD corrections reach
a minimum around $\Delta R_{\Pj_1 \Pj_2}=4$, while the EW corrections
tend to increase slightly towards increasing $\Delta R_{\Pj_1 \Pj_2}$,
resulting in an increase of the combined NLO prediction.  The
QCD corrections are generally flat and small for distributions in
leptonic angular variables, resulting in combined predictions very
close to the EW ones.  The distribution in the
rapidity--azimuthal-angle distance of the muon and anti-muon
(\reffi{plot:dR_mumu}) displays increasing EW corrections with
increasing $\Delta R_{\mu^-\mu^+}$, varying from $-45\%$ at $\Delta
R_{\mu^-\mu^+}\to0$ to about $-10\%$ at $\Delta R_{\mu^-\mu^+}=4$.
The distribution in the cosine of the angle
between the muon and anti-muon (\reffi{plot:cosphi_mumu}) receives
only a mild shape distortion towards
$\cos\theta_{\mu^-\mu^+}\to1$ from the EW corrections.

\begin{figure}
        \setlength{\parskip}{-10pt}
        \begin{subfigure}{0.49\textwidth}
                \subcaption{}
                 \includegraphics[width=\textwidth,page=8]{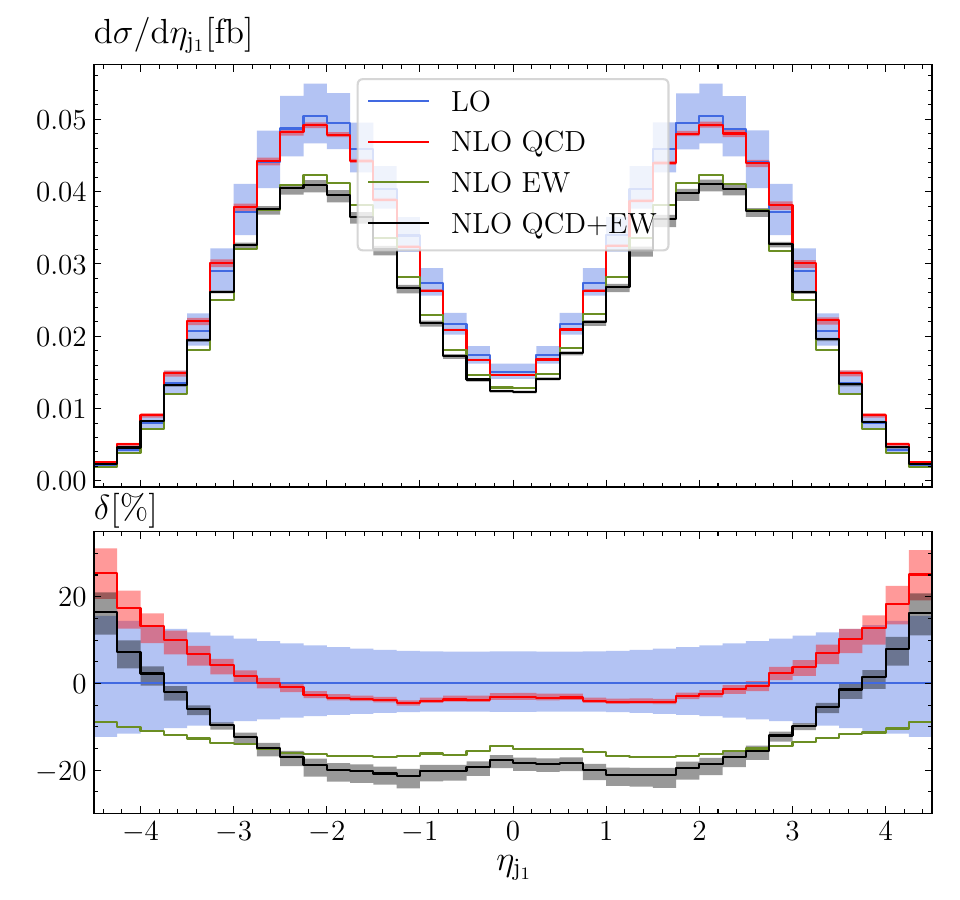}
                \label{plot:eta_j1}
        \end{subfigure}
        \hfill
        \begin{subfigure}{0.49\textwidth}
                \subcaption{}
                 \includegraphics[width=\textwidth,page=9]{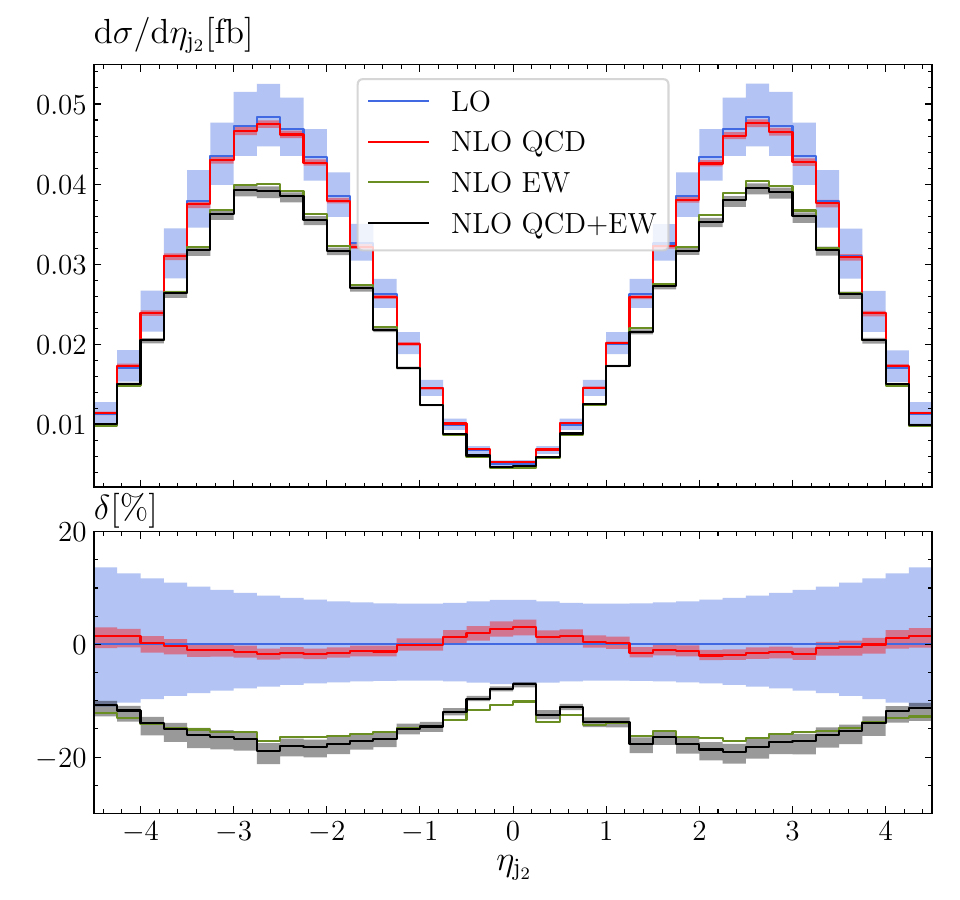}
                \label{plot:eta_j2}
        \end{subfigure}

        \begin{subfigure}{0.49\textwidth}
                \subcaption{}
                 \includegraphics[width=\textwidth,page=11]{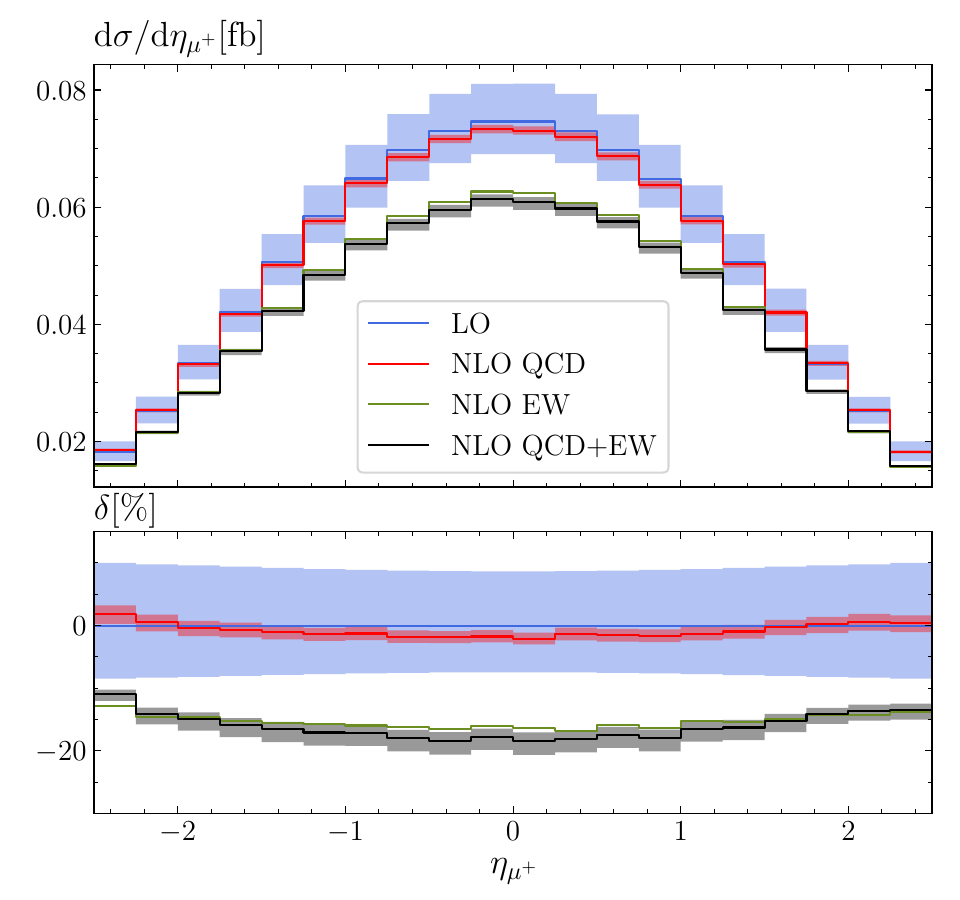}
                \label{plot:eta_mup}
        \end{subfigure}
        \hfill
        \begin{subfigure}{0.49\textwidth}
                \subcaption{}
                 \includegraphics[width=\textwidth,page=2]{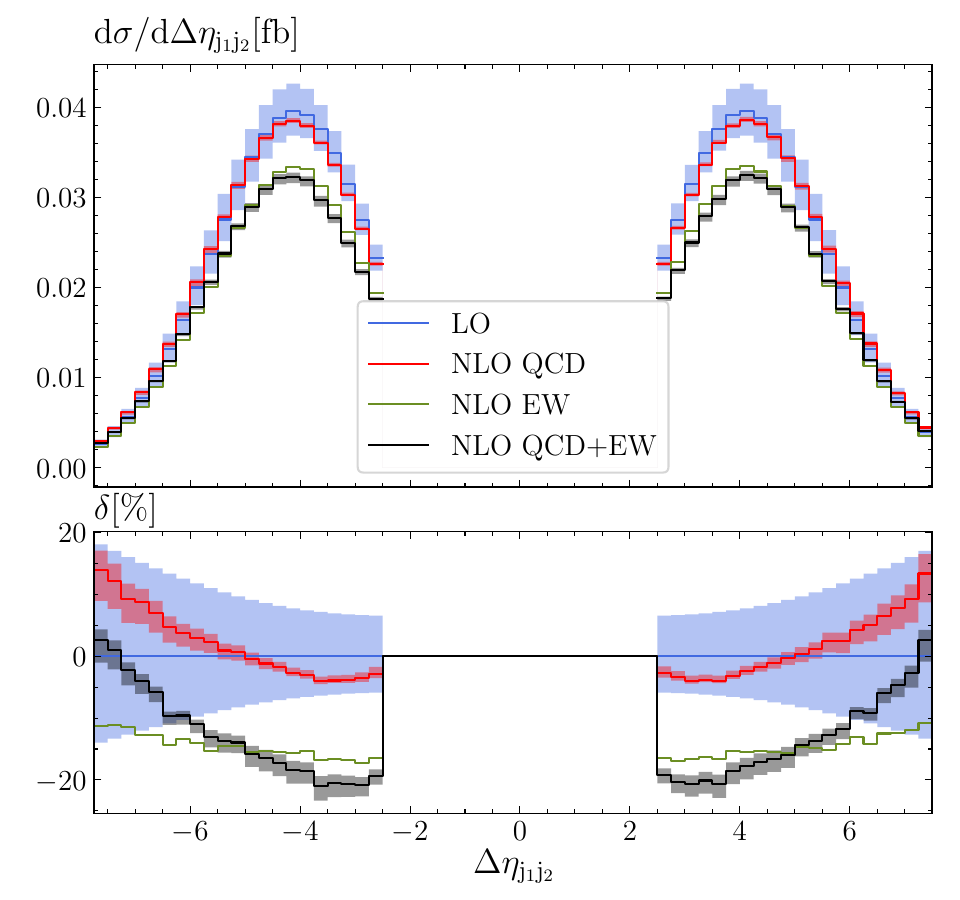}
                \label{plot:deta_j1j2}
        \end{subfigure}

        \vspace*{-3ex}
        \caption{\label{fig:rap}%
                Differential distributions for $\Pp\Pp \to \mu^+\mu^-\Pe^+\nu_\Pe\Pj\Pj+X$ at the LHC with CM energy $13\TeV$:
                \subref{plot:eta_j1}~pseudo-rapidity of the hardest jet~(top left), %
                \subref{plot:eta_j2}~pseudo-rapidity of the second
                hardest jet~(top right),
                \subref{plot:eta_mup}~pseudo-rapidity of the
                anti-muon~(bottom left), and
                \subref{plot:deta_j1j2}~difference of pseudo-rapidity of the
                two tagging jets~(bottom right).  The upper panel
                shows the LO contributions of order
                $\mathcal{O}{\left( \alpha^6 \right)}$, the two NLO
                  predictions [including $\mathcal{O}{\left(\alpha^7 \right)}$
                    (NLO EW) and $\mathcal{O}{\left( \alphas \alpha^6
                      \right)}$ (NLO QCD)] as well as their sum.  The
                    lower panel shows the relative NLO corrections
                    with respect to the LO in percent.}
\end{figure}
Figure~\ref{fig:rap} shows pseudo-rapidity distributions.
The first two concern the hardest (\reffi{plot:eta_j1}) and second hardest jet (\reffi{plot:eta_j2}).
Both QCD and EW corrections are rather similar
in shape and differ mainly by some offset.  
For the hardest jet, the corrections peak in the peripheral region, while for the
second hardest jet they increase in the central region as well.  The
distribution in the pseudo-rapidity of the anti-muon
(\reffi{plot:eta_mup}) receives flat corrections over almost the whole
range with only a slight increase in the peripheral region.
Finally, both QCD and EW corrections increase with growing
pseudo-rapidity difference of the two leading jets.

\begin{figure}
        \setlength{\parskip}{-10pt}
        \begin{subfigure}{0.49\textwidth}
                \subcaption{}
                 \includegraphics[width=\textwidth,page=14]{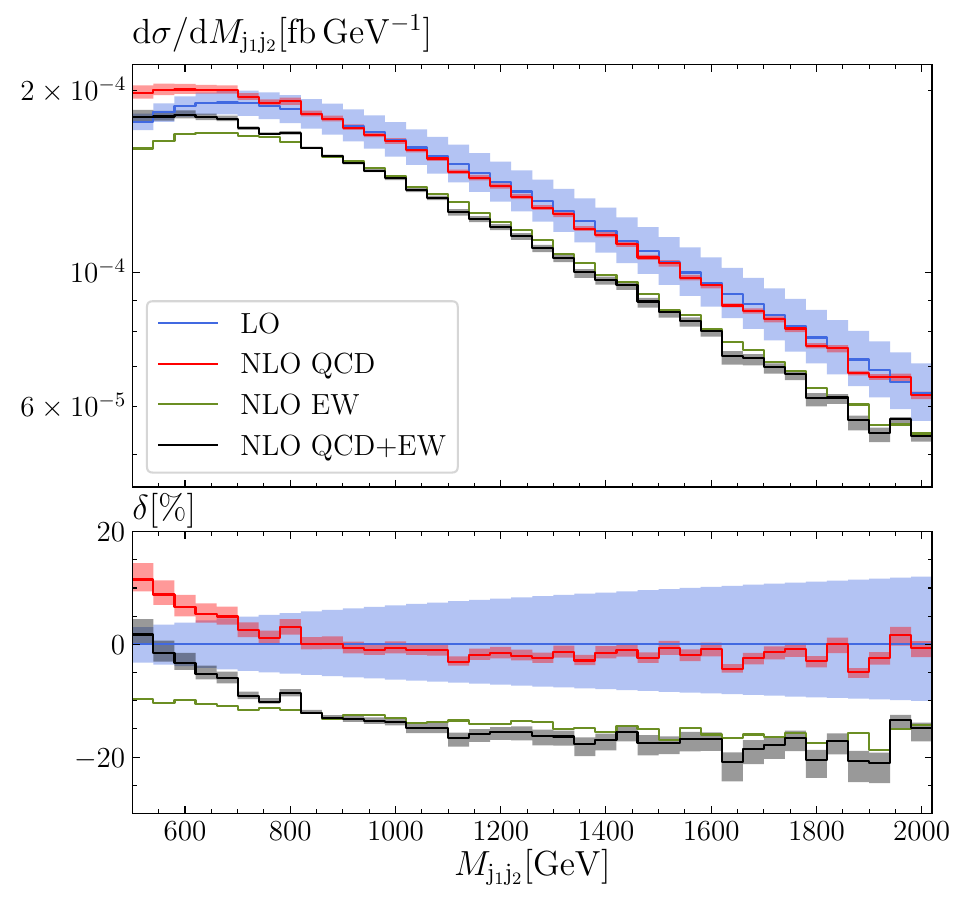}
                \label{plot:mjj}
        \end{subfigure}
        \hfill
        \begin{subfigure}{0.49\textwidth}
                \subcaption{}
                 \includegraphics[width=\textwidth,page=17]{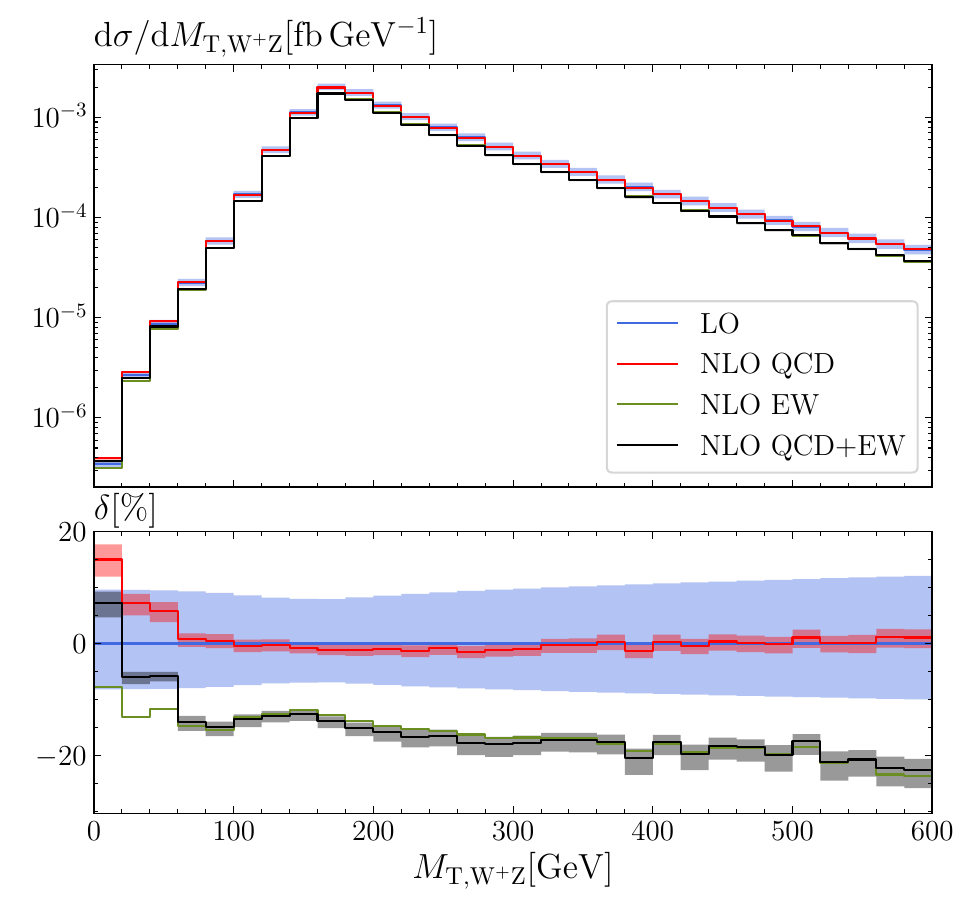}
                \label{plot:mTwz}
        \end{subfigure}

        \vspace*{-3ex}
        \caption{\label{fig:dist_m}%
          Differential distributions for $\Pp\Pp \to
          \mu^+\mu^-\Pe^+\nu_\Pe\Pj\Pj+X$ at the LHC with CM energy
          $13\TeV$: \subref{plot:mjj}~invariant mass of the two
          tagging jets~(left), and \subref{plot:mTwz}~transverse mass
          of the four lepton system~(right).  The upper panel shows
          the LO contributions of order $\mathcal{O}{\left( \alpha^6
            \right)}$, the two NLO predictions
            [including $\mathcal{O}{\left(\alpha^7 \right)}$ (NLO EW) and
            $\mathcal{O}{\left( \alphas \alpha^6 \right)}$ (NLO QCD)]
            as well as their sum.  The lower panel shows the relative
            NLO corrections with respect to the LO in percent.}
\end{figure}

Finally, we show the
distributions in the
invariant mass of the two jets (\reffi{plot:mjj}) and in the transverse
mass of the WZ system (\reffi{plot:mTwz}).  The invariant-mass
distribution of the two jets displays a similar behaviour than the one
in the transverse momentum of the jets: There is a steady increase in
size of the negative EW corrections towards more and more
negative values with increasing invariant masses due to EW high-energy
logarithms.  The QCD corrections are positive at $500\GeV$ and decrease slowly towards higher invariant masses. 
This is particularly interesting, as this observable is used to define fiducial regions in
measurements.  The transverse mass $M_{\mathrm{T,W^+ Z}}$, of the WZ
system is defined as
\begin{equation}
M^2_{\mathrm{T,W^+ Z}} =
{\Bigl( \sum_{\ell} \ptsub{\ell} \Bigr)^2- \Bigl(\sum_{\ell} p_{x,\ell}\Bigr)^2 - \Bigl(\sum_{\ell} p_{y,\ell}\Bigr)^2},
\end{equation}
where $\ell$ is running over the four leptons (including the
neutrino).  
In the region below $\MW+\MZ$, which does not receive 
contributions from doubly-resonant $\PW\PZ$~pairs, 
QCD and EW corrections are flat, while for large
transverse masses the Sudakov logarithms dominate the EW corrections.
The QCD corrections turn out to be small for the invariant-mass
distributions.

\section{Conclusion}
\label{sec:conclusion}

The process $\Pp\Pp \to \mu^+\mu^-\Pe^+\nu_\Pe\Pj\Pj+X$ is of great
interest at the LHC, because its EW contribution of order
$\mathcal{O}{\left(\alpha^6 \right)}$ to the cross section contains
vector-boson scattering (VBS) as a subprocess.  In this article we have
reported on a calculation of NLO corrections of order
$\mathcal{O}{\left( \alphas \alpha^6 \right)}$ and $\mathcal{O}{\left(
    \alpha^7 \right)}$ to the EW process.  This is the first time that
the EW corrections of $\mathcal{O}{\left( \alpha^7 \right)}$ are computed
for such a final state.  While the QCD corrections of order
$\mathcal{O}{\left( \alphas \alpha^6 \right)}$ have already been
computed in the VBS approximation \cite{Bozzi:2007ur}, for the first
time their full computation (including interference contributions
of EW type) is performed.  The combination of these two NLO contributions
constitutes the complete NLO prediction for the EW component of the
$\Pp\Pp \to \mu^+\mu^-\Pe^+\nu_\Pe\Pj\Pj+X$ process.

The EW corrections turn out to be relatively large, in accordance with
similar observations made already for like-sign W~scattering in
\citere{Biedermann:2016yds}.  This confirms the expectation that large
EW corrections are indeed an intrinsic feature of VBS at the LHC.  The
corrections reach $-16\%$ in the chosen fiducial region and are
driven by Sudakov logarithms that are large and negative and grow in
size in the high-energy limit.  The large EW corrections for the
integrated fiducial cross section can be well reproduced by a simple logarithmic
approximation. The corrections of order $\mathcal{O}{\left( \alphas
    \alpha^6 \right)}$ (sometimes simply called QCD corrections) are
small and negative.  They amount to about $-2\%$ at the level of the
integrated cross section and do not exceed $20\%$ in differential
distributions.

To complete the picture for the VBS process, we have also reported on
several LO contributions, including some suppressed channels.  The
interference contribution of order $\mathcal{O}{\left( \alphas
    \alpha^5 \right)}$ is below $1\%$, while the QCD contribution of
order $\mathcal{O}{\left( \alphas^2 \alpha^4 \right)}$ is larger than
the EW process by a factor 4.  This shows how challenging the
measurement of the EW process of order $\mathcal{O}{\left(\alpha^6
  \right)}$ is, highlighting therefore the need for precise
predictions in this context.  Moreover, we have computed all the LO
contributions with a photon in the initial state.  These turn out to
be rather small and can safely be neglected in future analyses.
Finally, we have computed the contributions involving bottom quarks (either
both in the initial state or one in the final and one in the initial
state) which turn out to be non-negligible.  These are enhanced by
contributions with a singly-resonant top quark, but can, in principle,
be experimentally suppressed using b-tagging techniques.

In addition to the phenomenological relevance of the presented
calculation, it is worth stressing that this also constitutes a
non-trivial extension of previous computations.  The complications
arising here are manifold: (i) This is the first time that a process
is computed at NLO EW accuracy with seven external charged particles.
This increases the complexity of the real contribution as well as 
of the virtual corrections.  (ii) The computation of the real contribution
entails additional complications.  QCD real radiation contains
singular contributions arising from soft/collinear photons and gluons.
This requires an advanced automation of computation of the real
corrections.  (iii) The number of partonic channels is very large.
With respect to the like-sign W channel, the number of partonic
channels increases by more than a factor three.  This means that the
computing time is considerably increased, and efficient book-keeping
and parallelisation become decisive.

To come as close as possible to the situation realised in experimental
analyses, our results are given in terms of integrated cross sections
and differential distributions in the so-called \emph{loose fiducial}
region presented in \citere{Sirunyan:2019ksz} by the CMS collaboration.
The event selections for this phase space are simplified with respect
to the ones used for the actual measurement.  Such efforts are
particularly welcome to theorists as they allow a direct use of state-of-the-art
theoretical progress in experimental analyses.  Therefore the
predictions provided in the present article should be particularly
useful for the VBS program at the LHC.

\section*{Acknowledgements}

AD and MP thank Jean-Nicolas Lang for
supporting {\sc Recola}.  
MP thanks Kenneth Long, Narei Lorenzo Martinez, Jakob Salfeld-Nebgen, and Marco Zaro for useful discussions.
AD and MP acknowledge financial support by
the German Federal Ministry for Education and Research (BMBF) under
contracts no.~05H15WWCA1 and 05H18WWCA1 and the German Research
Foundation (DFG) under reference number DE 623/6-1.  SD and CS
acknowledge support by the state of Baden-W\"urttemberg through bwHPC
and the DFG through grant no.\ INST 39/963-1 FUGG and grant DI 784/3.
MP is supported by the European Research Council
Consolidator Grant NNLOforLHC2.
CS is supported by the European Research Council under the European
Unions Horizon 2020 research and innovation Programme (grant agreement
no.\ 740006).
The authors would also like to acknowledge financial support from the COST Action CA16108.

\bibliographystyle{JHEPmod}
\bibliography{vbs_wz}

\providecommand{\href}[2]{#2}\begingroup\raggedright\begin{thebibliography}{10}

\bibitem{Aad:2014zda}
{\bf ATLAS} Collaboration, G.~Aad et~al., {\it {Evidence for Electroweak
  Production of $W^{\pm}W^{\pm}jj$ in $pp$ Collisions at $\sqrt{s}={}$8\,TeV
  with the ATLAS Detector}},  {\em Phys. Rev. Lett.} {\bf 113} (2014) 141803,
  [\href{http://arxiv.org/abs/1405.6241}{{\tt arXiv:1405.6241}}].

\bibitem{Khachatryan:2014sta}
{\bf CMS} Collaboration, V.~Khachatryan et~al., {\it {Study of vector boson
  scattering and search for new physics in events with two same-sign leptons
  and two jets}},  {\em Phys. Rev. Lett.} {\bf 114} (2015) 051801,
  [\href{http://arxiv.org/abs/1410.6315}{{\tt arXiv:1410.6315}}].

\bibitem{Aaboud:2016ffv}
{\bf ATLAS} Collaboration, M.~Aaboud et~al., {\it {Measurement of
  $W^{\pm}W^{\pm}$ vector-boson scattering and limits on anomalous quartic
  gauge couplings with the ATLAS detector}},  {\em Phys. Rev.} {\bf D96} (2017)
  012007, [\href{http://arxiv.org/abs/1611.02428}{{\tt arXiv:1611.02428}}].

\bibitem{Sirunyan:2017ret}
{\bf CMS} Collaboration, A.~M. Sirunyan et~al., {\it {Observation of
  electroweak production of same-sign W boson pairs in the two jet and two
  same-sign lepton final state in proton-proton collisions at $\sqrt{s}
  ={}$13\,TeV}},  {\em Phys. Rev. Lett.} {\bf 120} (2018) 081801,
  [\href{http://arxiv.org/abs/1709.05822}{{\tt arXiv:1709.05822}}].

\bibitem{ATLAS-CONF-2018-030}
{\bf ATLAS} Collaboration, ``{Observation of electroweak production of a
  same-sign $W$ boson pair in association with two jets in $pp$ collisions at
  $\sqrt{s}=$13\,TeV with the ATLAS detector}.'' ATLAS-CONF-2018-030, 2018.

\bibitem{ATLAS:2018ucv}
{\bf ATLAS} Collaboration, ``{Observation of electroweak $W^{\pm}Z$ boson pair
  production in association with two jets in pp collisions at
  $\sqrt{s}={}$13\,TeV with the ATLAS Detector}.'' ATLAS-CONF-2018-033, 2018.

\bibitem{Sirunyan:2019ksz}
{\bf CMS} Collaboration, A.~M. Sirunyan et~al., {\it {Measurement of
  electroweak WZ boson production and search for new physics in WZ $+$ two jets
  events in pp collisions at $\sqrt{s} =$ 13 TeV}},
  \href{http://arxiv.org/abs/1901.04060}{{\tt arXiv:1901.04060}}.

\bibitem{Biedermann:2016yds}
B.~Biedermann, A.~Denner, and M.~Pellen, {\it {Large electroweak corrections to
  vector-boson scattering at the Large Hadron Collider}},  {\em Phys. Rev.
  Lett.} {\bf 118} (2017) 261801, [\href{http://arxiv.org/abs/1611.02951}{{\tt
  arXiv:1611.02951}}].

\bibitem{Biedermann:2017bss}
B.~Biedermann, A.~Denner, and M.~Pellen, {\it {Complete NLO corrections to
  W$^{+}$W$^{+}$ scattering and its irreducible background at the LHC}},  {\em
  JHEP} {\bf 10} (2017) 124, [\href{http://arxiv.org/abs/1708.00268}{{\tt
  arXiv:1708.00268}}].

\bibitem{Bendavid:2018nar}
J.~R. Andersen et~al., {\it {Les Houches 2017: Physics at TeV Colliders
  Standard Model Working Group Report}},  in {\em {10th Les Houches Workshop on
  Physics at TeV Colliders (PhysTeV 2017) Les Houches, France, June 5-23,
  2017}}, 2018.
\newblock \href{http://arxiv.org/abs/1803.07977}{{\tt arXiv:1803.07977}}.

\bibitem{Bozzi:2007ur}
G.~Bozzi, B.~{J\"ager}, C.~Oleari, and D.~Zeppenfeld, {\it {Next-to-leading
  order QCD corrections to $W^+ Z$ and $W^- Z$ production via vector-boson
  fusion}},  {\em Phys. Rev.} {\bf D75} (2007) 073004,
  [\href{http://arxiv.org/abs/hep-ph/0701105}{{\tt hep-ph/0701105}}].

\bibitem{Jager:2018cyo}
B.~Jäger, A.~Karlberg, and J.~Scheller, {\it {Parton-shower effects in
  electroweak $WZjj$ production at the next-to-leading order of QCD}},  {\em
  Eur. Phys. J.} {\bf C79} (2019) 226,
  [\href{http://arxiv.org/abs/1812.05118}{{\tt arXiv:1812.05118}}].

\bibitem{Ballestrero:2018anz}
A.~Ballestrero et~al., {\it {Precise predictions for same-sign W-boson
  scattering at the LHC}},  {\em Eur. Phys. J.} {\bf C78} (2018) 671,
  [\href{http://arxiv.org/abs/1803.07943}{{\tt arXiv:1803.07943}}].

\bibitem{Cascioli:2011va}
F.~Cascioli, P.~Maierh{\"o}fer, and S.~Pozzorini, {\it {Scattering Amplitudes
  with Open Loops}},  {\em Phys. Rev. Lett.} {\bf 108} (2012) 111601,
  [\href{http://arxiv.org/abs/1111.5206}{{\tt arXiv:1111.5206}}].

\bibitem{Kallweit:2014xda}
S.~Kallweit, J.~M. Lindert, P.~Maierh{\"o}fer, S.~Pozzorini, and
  M.~Sch{\"o}nherr, {\it {NLO electroweak automation and precise predictions
  for W+multijet production at the LHC}},  {\em JHEP} {\bf 04} (2015) 012,
  [\href{http://arxiv.org/abs/1412.5157}{{\tt arXiv:1412.5157}}].

\bibitem{Denner:2014gla}
A.~Denner, S.~Dittmaier, and L.~Hofer, {\it {COLLIER - A fortran-library for
  one-loop integrals}},  {\em PoS} {\bf LL2014} (2014) 071,
  [\href{http://arxiv.org/abs/1407.0087}{{\tt arXiv:1407.0087}}].

\bibitem{Denner:2016kdg}
A.~Denner, S.~Dittmaier, and L.~Hofer, {\it {COLLIER: a fortran-based Complex
  One-Loop LIbrary in Extended Regularizations}},  {\em Comput. Phys. Commun.}
  {\bf 212} (2017) 220--238, [\href{http://arxiv.org/abs/1604.06792}{{\tt
  arXiv:1604.06792}}].

\bibitem{Actis:2012qn}
S.~Actis, A.~Denner, L.~Hofer, A.~Scharf, and S.~Uccirati, {\it {Recursive
  generation of one-loop amplitudes in the Standard Model}},  {\em JHEP} {\bf
  04} (2013) 037, [\href{http://arxiv.org/abs/1211.6316}{{\tt
  arXiv:1211.6316}}].

\bibitem{Actis:2016mpe}
S.~Actis, et~al., {\it {RECOLA: REcursive Computation of One-Loop Amplitudes}},
   {\em Comput. Phys. Commun.} {\bf 214} (2017) 140--173,
  [\href{http://arxiv.org/abs/1605.01090}{{\tt arXiv:1605.01090}}].

\bibitem{Dittmaier:1999mb}
S.~Dittmaier, {\it {A general approach to photon radiation off fermions}},
  {\em Nucl. Phys.} {\bf B565} (2000) 69--122,
  [\href{http://arxiv.org/abs/hep-ph/9904440}{{\tt hep-ph/9904440}}].

\bibitem{Dittmaier:2008md}
S.~Dittmaier, A.~Kabelschacht, and T.~Kasprzik, {\it {Polarized QED splittings
  of massive fermions and dipole subtraction for non-collinear-safe
  observables}},  {\em Nucl. Phys.} {\bf B800} (2008) 146--189,
  [\href{http://arxiv.org/abs/0802.1405}{{\tt arXiv:0802.1405}}].

\bibitem{xxx}
A.~Denner, S.~Dittmaier, M.~Pellen, and C.~Schwan, ``{Low-virtuality photon
  transitions $\gamma^*\to f\bar f$ and the photon-to-jet conversion
  function}.'' In preparation.

\bibitem{hep-mc}
``{A C++11 Template Library for Monte Carlo Integration}.''
  https://github.com/cschwan/hep-mc.

\bibitem{Dittmaier:2002ap}
S.~Dittmaier and M.~Roth, {\it {LUSIFER: A LUcid approach to six FERmion
  production}},  {\em Nucl. Phys.} {\bf B642} (2002) 307--343,
  [\href{http://arxiv.org/abs/hep-ph/0206070}{{\tt hep-ph/0206070}}].

\bibitem{MpiForum}
``{MPI Forum}.'' https://www.mpi-forum.org/.

\bibitem{Denner:1999gp}
A.~Denner, S.~Dittmaier, M.~Roth, and D.~Wackeroth, {\it {Predictions for all
  processes ${e}^+ {e}^- \to $ 4 fermions $+ \gamma$}},  {\em Nucl. Phys.} {\bf
  B560} (1999) 33--65, [\href{http://arxiv.org/abs/hep-ph/9904472}{{\tt
  hep-ph/9904472}}].

\bibitem{Berends:1994pv}
F.~A. Berends, R.~Pittau, and R.~Kleiss, {\it {All electroweak four fermion
  processes in electron-positron collisions}},  {\em Nucl. Phys.} {\bf B424}
  (1994) 308--342, [\href{http://arxiv.org/abs/hep-ph/9404313}{{\tt
  hep-ph/9404313}}].

\bibitem{Denner:2015yca}
A.~Denner and R.~Feger, {\it {NLO QCD corrections to off-shell top-antitop
  production with leptonic decays in association with a Higgs boson at the
  LHC}},  {\em JHEP} {\bf 11} (2015) 209,
  [\href{http://arxiv.org/abs/1506.07448}{{\tt arXiv:1506.07448}}].

\bibitem{Denner:2016jyo}
A.~Denner and M.~Pellen, {\it {NLO electroweak corrections to off-shell
  top-antitop production with leptonic decays at the LHC}},  {\em JHEP} {\bf
  08} (2016) 155, [\href{http://arxiv.org/abs/1607.05571}{{\tt
  arXiv:1607.05571}}].

\bibitem{Denner:2016wet}
A.~Denner, J.-N. Lang, M.~Pellen, and S.~Uccirati, {\it {Higgs production in
  association with off-shell top-antitop pairs at NLO EW and QCD at the LHC}},
  {\em JHEP} {\bf 02} (2017) 053, [\href{http://arxiv.org/abs/1612.07138}{{\tt
  arXiv:1612.07138}}].

\bibitem{Denner:2017kzu}
A.~Denner and M.~Pellen, {\it {Off-shell production of top-antitop pairs in the
  lepton+jets channel at NLO QCD}},  {\em JHEP} {\bf 02} (2018) 013,
  [\href{http://arxiv.org/abs/1711.10359}{{\tt arXiv:1711.10359}}].

\bibitem{Catani:1996vz}
S.~Catani and M.~H. Seymour, {\it {A general algorithm for calculating jet
  cross-sections in NLO QCD}},  {\em Nucl. Phys.} {\bf B485} (1997) 291--419,
  [\href{http://arxiv.org/abs/hep-ph/9605323}{{\tt hep-ph/9605323}}]. [Erratum:
  Nucl. Phys. {\bf B510} (1998) 503].

\bibitem{Buckley:2014ana}
A.~Buckley, et~al., {\it {LHAPDF6: parton density access in the LHC precision
  era}},  {\em Eur. Phys. J.} {\bf C75} (2015) 132,
  [\href{http://arxiv.org/abs/1412.7420}{{\tt arXiv:1412.7420}}].

\bibitem{'tHooft:1978xw}
G.~'t~Hooft and M.~J.~G. Veltman, {\it {Scalar one loop integrals}},  {\em
  Nucl. Phys.} {\bf B153} (1979) 365--401.

\bibitem{Beenakker:1988jr}
W.~Beenakker and A.~Denner, {\it {Infrared divergent scalar box integrals with
  applications in the electroweak Standard Model}},  {\em Nucl. Phys.} {\bf
  B338} (1990) 349--370.

\bibitem{Dittmaier:2003bc}
S.~Dittmaier, {\it {Separation of soft and collinear singularities from
  one-loop N point integrals}},  {\em Nucl. Phys.} {\bf B675} (2003) 447--466,
  [\href{http://arxiv.org/abs/hep-ph/0308246}{{\tt hep-ph/0308246}}].

\bibitem{Denner:2010tr}
A.~Denner and S.~Dittmaier, {\it {Scalar one-loop 4-point integrals}},  {\em
  Nucl. Phys.} {\bf B844} (2011) 199--242,
  [\href{http://arxiv.org/abs/1005.2076}{{\tt arXiv:1005.2076}}].

\bibitem{Passarino:1978jh}
G.~Passarino and M.~J.~G. Veltman, {\it {One-loop corrections for ${e}^+ {e}^-$
  annihilation into $\mu^+\mu^-$ in the Weinberg Model}},  {\em Nucl. Phys.}
  {\bf B160} (1979) 151.

\bibitem{Denner:2002ii}
A.~Denner and S.~Dittmaier, {\it {Reduction of one-loop tensor 5-point
  integrals}},  {\em Nucl. Phys.} {\bf B658} (2003) 175--202,
  [\href{http://arxiv.org/abs/hep-ph/0212259}{{\tt hep-ph/0212259}}].

\bibitem{Denner:2005nn}
A.~Denner and S.~Dittmaier, {\it {Reduction schemes for one-loop tensor
  integrals}},  {\em Nucl. Phys.} {\bf B734} (2006) 62--115,
  [\href{http://arxiv.org/abs/hep-ph/0509141}{{\tt hep-ph/0509141}}].

\bibitem{Denner:2005fg}
A.~Denner, S.~Dittmaier, M.~Roth, and L.~H. Wieders, {\it {Electroweak
  corrections to charged-current ${e}^+ {e}^- \to$ 4 fermion processes:
  Technical details and further results}},  {\em Nucl. Phys.} {\bf B724} (2005)
  247--294, [\href{http://arxiv.org/abs/hep-ph/0505042}{{\tt hep-ph/0505042}}].
  [Erratum: Nucl. Phys. {\bf B854} (2012) 504].

\bibitem{Denner:2006ic}
A.~Denner and S.~Dittmaier, {\it {The complex-mass scheme for perturbative
  calculations with unstable particles}},  {\em Nucl. Phys. Proc. Suppl.} {\bf
  160} (2006) 22--26, [\href{http://arxiv.org/abs/hep-ph/0605312}{{\tt
  hep-ph/0605312}}].

\bibitem{Nagy:1998bb}
Z.~Nagy and Z.~Trócsányi, {\it {Next-to-leading order calculation of four-jet
  observables in electron-positron annihilation}},  {\em Phys. Rev.} {\bf D59}
  (1999) 014020, [\href{http://arxiv.org/abs/hep-ph/9806317}{{\tt
  hep-ph/9806317}}]. [Erratum: Phys. Rev. {\bf D62} (2000) 099902].

\bibitem{Ball:2014uwa}
{\bf NNPDF} Collaboration, R.~D. Ball et~al., {\it {Parton distributions for
  the LHC Run II}},  {\em JHEP} {\bf 04} (2015) 040,
  [\href{http://arxiv.org/abs/1410.8849}{{\tt arXiv:1410.8849}}].

\bibitem{Bertone:2017bme}
{\bf NNPDF} Collaboration, V.~Bertone, S.~Carrazza, N.~P. Hartland, and
  J.~Rojo, {\it {Illuminating the photon content of the proton within a global
  PDF analysis}},  {\em SciPost Phys.} {\bf 5} (2018) 008,
  [\href{http://arxiv.org/abs/1712.07053}{{\tt arXiv:1712.07053}}].

\bibitem{Manohar:2016nzj}
A.~Manohar, P.~Nason, G.~P. Salam, and G.~Zanderighi, {\it {How bright is the
  proton? A precise determination of the photon parton distribution function}},
   {\em Phys. Rev. Lett.} {\bf 117} (2016) 242002,
  [\href{http://arxiv.org/abs/1607.04266}{{\tt arXiv:1607.04266}}].

\bibitem{Manohar:2017eqh}
A.~V. Manohar, P.~Nason, G.~P. Salam, and G.~Zanderighi, {\it {The Photon
  Content of the Proton}},  {\em JHEP} {\bf 12} (2017) 046,
  [\href{http://arxiv.org/abs/1708.01256}{{\tt arXiv:1708.01256}}].

\bibitem{Denner:2012dz}
A.~Denner, L.~Ho\v{s}ekov\'a, and S.~Kallweit, {\it {NLO QCD corrections to
  $W^+ W^+ jj$ production in vector-boson fusion at the LHC}},  {\em Phys.
  Rev.} {\bf D86} (2012) 114014, [\href{http://arxiv.org/abs/1209.2389}{{\tt
  arXiv:1209.2389}}].

\bibitem{Denner:2000bj}
A.~Denner, S.~Dittmaier, M.~Roth, and D.~Wackeroth, {\it {Electroweak radiative
  corrections to ${e}^+ {e}^- \to {W W} \to$ 4 fermions in double pole
  approximation: The RACOONWW approach}},  {\em Nucl. Phys.} {\bf B587} (2000)
  67--117, [\href{http://arxiv.org/abs/hep-ph/0006307}{{\tt hep-ph/0006307}}].

\bibitem{Dittmaier:2001ay}
S.~Dittmaier and M.~Kr{\"a}mer, {\it {Electroweak radiative corrections to
  W-boson production at hadron colliders}},  {\em Phys. Rev.} {\bf D65} (2002)
  073007, [\href{http://arxiv.org/abs/hep-ph/0109062}{{\tt hep-ph/0109062}}].

\bibitem{Tanabashi:2018oca}
{\bf ParticleDataGroup} Collaboration, M.~Tanabashi et~al., {\it {Review of
  Particle Physics}},  {\em Phys. Rev.} {\bf D98} (2018) 030001.

\bibitem{Basso:2015gca}
L.~Basso, S.~Dittmaier, A.~Huss, and L.~Oggero, {\it {Techniques for the
  treatment of IR divergences in decay processes at NLO and application to the
  top-quark decay}},  {\em Eur. Phys. J.} {\bf C76} (2016) 56,
  [\href{http://arxiv.org/abs/1507.04676}{{\tt arXiv:1507.04676}}].

\bibitem{Heinemeyer:2013tqa}
{\bf LHC Higgs Cross Section Working Group} Collaboration, J.~R. Andersen
  et~al., ``{Handbook of LHC Higgs Cross Sections: 3. Higgs Properties}.''
  CERN-2013-004, FERMILAB-CONF-13-667-T, 2013.

\bibitem{Bardin:1988xt}
D.~{\relax Yu}. Bardin, A.~Leike, T.~Riemann, and M.~Sachwitz, {\it
  {Energy-dependent width effects in ${e}^+ {e}^-$-annihilation near the
  Z-boson pole}},  {\em Phys. Lett.} {\bf B206} (1988) 539--542.

\bibitem{Cacciari:2008gp}
M.~Cacciari, G.~P. Salam, and G.~Soyez, {\it {The anti-$k_t$ jet clustering
  algorithm}},  {\em JHEP} {\bf 04} (2008) 063,
  [\href{http://arxiv.org/abs/0802.1189}{{\tt arXiv:0802.1189}}].

\bibitem{Sirunyan:2018zgs}
{\bf CMS} Collaboration, A.~M. Sirunyan et~al., {\it {Observation of single top
  quark production in association with a Z boson in proton-proton collisions at
  $\sqrt{s} =$ 13 TeV}},  {\em Phys. Rev. Lett.} {\bf 122} (2019) 132003,
  [\href{http://arxiv.org/abs/1812.05900}{{\tt arXiv:1812.05900}}].

\bibitem{Denner:2000jv}
A.~Denner and S.~Pozzorini, {\it {One loop leading logarithms in electroweak
  radiative corrections. 1. Results}},  {\em Eur. Phys. J.} {\bf C18} (2001)
  461--480, [\href{http://arxiv.org/abs/hep-ph/0010201}{{\tt hep-ph/0010201}}].

\bibitem{Denner:1997kq}
A.~Denner and T.~Hahn, {\it {Radiative corrections to $W^+ W^-\to W^+ W^-$ in
  the electroweak standard model}},  {\em Nucl. Phys.} {\bf B525} (1998)
  27--50, [\href{http://arxiv.org/abs/hep-ph/9711302}{{\tt hep-ph/9711302}}].

\bibitem{Azzi:2019yne}
{\bf HL-LHC and HE-LHC Working Group} Collaboration, P.~Azzi et~al., {\it
  {Standard Model Physics at the HL-LHC and HE-LHC}},
  \href{http://arxiv.org/abs/1902.04070}{{\tt arXiv:1902.04070}}.

\end{thebibliography}\endgroup

\end{document}